\documentclass[useAMS,usenatbib]{mn2e}
\usepackage[english]{babel}

\usepackage{fancyhdr}
\usepackage{graphicx}
\usepackage{times}
\usepackage{natbib}
\usepackage{amsmath}
\usepackage{amsfonts}
\usepackage{amssymb}
\usepackage{multicol}
\usepackage{layout}
\usepackage{multirow}

\newif\ifAMStwofonts
\AMStwofontstrue

\graphicspath{{./fig/}}

\title[Shear and rotation for clustering dark energy]{Effects of shear and rotation on the spherical collapse model 
for clustering dark energy}

\author[F. Pace et~al.]{Francesco~Pace$^{1}$\thanks{E-mail: Francesco.Pace@port.ac.uk}, Ronaldo~C.~Batista$^{2}$,
Antonino~Del~Popolo$^{3,4}$\\
${}^1$ Institute for Cosmology and Gravitation, University of Portsmouth, Dennis Sciama Building, Portsmuth PO1 3FX, 
UK\\
${}^2$ Escola de Ci\^encias e Tecnologia, Universidade Federal do Rio Grande do Norte\\
Caixa Postal 1524, 59072-970, Natal, Rio Grande do Norte, Brazil\\
${}^3$ Dipartimento di Fisica e Astronomia, University Of Catania, Viale Andrea Doria 6, 95125 Catania, Italy\\
${}^4$ International Institute of Physics, Universidade Federal do Rio Grande do Norte, 59012-970 Natal, Brazil}

\begin{document}

\date{Accepted ?, Received ?; in original form ?}

\pagerange{\pageref{firstpage}--\pageref{lastpage}}\pubyear{0}

\maketitle

\label{firstpage}

\begin{abstract}
In the framework of the spherical collapse model we study the influence of shear and rotation terms for dark 
matter fluid in clustering dark energy models. We evaluate, for different equations of state, the effects of these 
terms on the linear overdensity threshold parameter, $\delta_{\rm c}$, and on the virial overdensity, 
$\Delta_{\rm V}$. The evaluation of their effects on $\delta_{\rm c}$ allows us to infer the modifications occurring 
on the mass function. Due to ambiguities in the definition of the halo mass in the case of clustering dark energy, we 
consider two different situations: the first is the classical one where the mass is of the dark matter halo only, 
while the second one is given by the sum of the mass of dark matter and dark energy. As previously found, the 
spherical collapse model becomes mass dependant and the two additional terms oppose to the collapse of the 
perturbations, especially on galactic scales, with respect to the spherical non-rotating model, while on clusters 
scales the effects of shear and rotation become negligible. The values for $\delta_{\rm c}$ and 
$\Delta_{\rm V}$ are higher than the standard spherical model. Regarding the effects of the additional non-linear 
terms on the mass function, we evaluate the number density of halos. As expected, major differences appear at high 
masses and redshifts. In particular, quintessence (phantom) models predict more (less) objects with respect to the 
$\Lambda$CDM model and the mass correction due to the contribution of the dark energy component has negligible effects 
on the overall number of structures.
\end{abstract}


\section{Introduction}
One of the most complex puzzle in modern cosmology is the understanding of the nature of the accelerated expansion of 
the Universe. This astonishing fact is the result of observations of high-redshifts supernovae, that are less luminous 
of what was expected in a decelerated universe \citep{Riess1998,Perlmutter1999,Tonry2003}. Assuming General Relativity 
and interpreting the dimming of Type Ia supernovae as due to an accelerated expansion phase in the history of the 
Universe, we are forced to introduce a new component with negative pressure, and in particular, to cause accelerated 
expansion, its equation-of-state parameter must be $w<-1/3$. This fluid, usually dubbed {\it dark energy} (DE), is 
totally unknown in its nature and physical characteristics.

The latest observations of Supernovae type Ia
\citep{Riess1998,Perlmutter1999,Knop2003,Riess2004,Astier2006,Riess2007,Amanullah2010}, 
together with the cosmic microwave background (CMB)
\citep{Komatsu2011,Jarosik2011,Planck2013_XV,Planck2013_XVI,Planck2013_XIX,Sievers2013}, 
the integrated Sachs-Wolfe effect (ISW) \citep{Giannantonio2008,Ho2008}, 
the large scale structure (LSS) and baryonic acoustic oscillations
\citep{Tegmark2004a,Tegmark2004b,Cole2005,Eisenstein2005,Percival2010,Reid2010,Blake2011a}, the globular clusters 
\citep{Krauss2003,Dotter2011}, high redshift galaxies \citep{Alcaniz2003} and the galaxy clusters 
\citep{Haiman2001,Allen2004,Allen2008,Wang2004,Basilakos2010} till works based on weak gravitational lensing 
\citep{Hoekstra2006,Jarvis2006} and X-ray clusters \citep{Vikhlinin2009} confirmed these early findings and they are 
all in agreement with a universe filled with 30\% by cold dark matter and baryons (both fluids pressureless) and with  
the remaining 70\% by the cosmological constant $\Lambda$ (the so-called $\Lambda$CDM model). The cosmological 
constant is the most basic form of dark energy. Its equation of state is constant in time ($w=-1$), it appears in 
Einstein field equations as a geometrical term, it cannot cluster (being constant in space and time) and its 
importance is appreciable only at low redshift.

Despite being in agreement virtually with all the observables, the standard cosmological model suffers of severe 
theoretical problems (the coincidence and the fine tuning problems) and therefore alternative models have been 
explored \citep[but see also][]{Astashenok2012}. The most studied ones are minimally coupled scalar fields 
(quintessence models). Since gravity is the main interaction acting on large scales, it is commonly believed that 
structures in the Universe formed via gravitational instability of primordial overdense perturbations that originated 
in the primeval inflationary phase \cite[][]{Starobinsky1980,Guth1981,Linde1990} of the Universe from quantum, 
Gaussian distributed fluctuations 
\citep[][]{DelPopolo2007,Komatsu2010,Casaponsa2011,Curto2011,Komatsu2011,Hinshaw2013,DelPopolo2014}.

Differently from the cosmological constant, even if often neglected in literature, dynamical dark energy models 
posses fluctuations that can alter the evolution of structure formation, not only via slowing down the growth rate 
but also giving rise to DE overdensities and underdensities which can evolve into the non-linear regime.

To study structure formation in the highly non-linear regime, it is very useful to work within the framework of the 
spherical collapse model, introduced by \cite{Gunn1972} and extended in many following works 
\citep[][]{Fillmore1984,Bertschinger1985,Hoffman1985,Ryden1987,Avila-Reese1998,Subramanian2000,Ascasibar2004,Mota2004, 
Williams2004,Abramo2007,Pace2010,Pace2014}. According to the model, perturbations are considered as being spherically 
symmetric non-rotating objects that, being overdense, decouple from the background Hubble expansion, reach a point of 
maximum expansion (turn-around) and collapse (formally to a singularity). In reality this does not happen and the 
kinetic energy associated with the collapse is converted into random motions creating an equilibrium configuration (a 
virialized structure).

Despite its crude approximations, the model is very successful in reproducing results of N-body simulations when 
combined with mass the function formalism \citep[][]{DelPopolo2007,Hiotelis2013}, either in usual minimally coupled 
dark energy models \citep{Pace2010} or in non-minimally coupled dark energy models \citep{Pace2014}. Nevertheless it 
is important to extend the basic formalism to include additional terms and make it more realistic.\\
Consequences of relaxing the sphericity assumption were studied by \cite{Hoffman1986,Hoffman1989} and 
\cite{Zaroubi1993}, while the introduction of radial motions and angular momentum was deeply studied by 
\cite{Ryden1987} and \cite{Gurevich1988a,Gurevich1988b}. We refer to \cite{DelPopolo2013b} for a more complete list of 
references and for details on the different models and how to link the angular momentum to the matter overdensity.

In this work we will extend previous works on the extended spherical collapse model in dark energy models 
\citep{DelPopolo2013a,DelPopolo2013b,DelPopolo2013c} by taking into account perturbations of the DE fluid. Since there 
are no N-body simulations with clustering dark energy so far, such study is valuable in order to have an idea about 
how dark energy fluctuations impact structure formation in a more realistic scenario. By writing the differential 
equations describing the dynamics of dark matter and dark energy, we will show how to relate the additional terms 
(shear and angular momentum) to the overdensity field and we will solve them to derive the time evolution of the 
typical parameters of the spherical collapse model, in particular the linear overdensity threshold for collapse 
$\delta_{\rm c}$ and the virial overdensity $\Delta_{\rm V}$ and we will show how these quantities are modified by the 
introduction of non zero vorticity and shear terms. Afterwards we will show how the mass function and its 
phenomenological extension to include DE perturbations are affected.

The paper is organised as following. In Section~\ref{sect:models} we discuss and summarise the dark energy models used 
in this work. In Section~\ref{sect:escm} we briefly derive the equations of the extended spherical collapse model 
whose solution will lead to the evaluation of $\delta_{\rm c}$ and $\Delta_{\rm V}$ (see 
Section~\ref{sect:escm_params}). In Section~\ref{sect:results} we show our results and in particular we devote 
Section~\ref{sect:mf} to the discussion of the effects of shear and rotation on the mass function. Finally, we 
conclude in Section~\ref{sect:conclusions}.

\section{The models}\label{sect:models}
For this work we use dark energy models previously analysed in the framework of the spherical collapse model where 
the usual assumption of negligible dark energy fluctuations is relaxed.\\
Dark energy models, described by an equation of state $w=P/(\rho c^2)$, either constant or time dependant, satisfy the 
background continuity equation
\begin{equation}\label{eqn:deceq}
 \dot{\rho}+3H(1+w)\rho=0\;.
\end{equation}
We consider eight different models and for the ones characterised by an evolving equation-of-state parameter we 
adopted the \cite{Chevallier2001} and \cite{Linder2003} (CPL) linear parametrization
\begin{equation}
 w(a)=w_0+(1-a)w_a\;,
\end{equation}
where $w_0$ and $w_a$ are constants and $a$ is the scale factor.

The reference model is the standard $\Lambda$CDM model where dark energy is represented by the cosmological constant 
with equation-of-state parameter $w=-1$, constant along the whole cosmic history. A consequence of this 
parametrization is that at early times this model behaves essentially as the EdS model (with $\Omega_{\rm m}=1$ and 
$\Omega_{\rm de}=0$) and the influence of the cosmological constant becomes appreciable only late in the cosmic 
history.
 
Due to the latest observational results by the {\it Planck} satellite\footnote{http://sci.esa.int/planck/} 
\citep{Planck2013_XV,Planck2013_XVI,Planck2013_XXVI}, we will assume a spatially-flat model.

Of the remaining seven models, two have a constant equation-of-state parameter $w>-1$ (the quintessence models DE1 
and DE2) and they differ from each other solely for the exact value of $w$. Other two instead have $w<-1$ (the phantom 
models DE3 and DE4). The latter are justified by recent Supernovae Type Ia (SNIa) observations 
\citep{Novosyadlyj2012,Planck2013_XVI,Rest2013,Scolnic2013,Shafer2014}.\\
Finally we consider three additional models with a time varying equation of state. Once again we can distinguish them 
according to the general behaviour of the equation-of-state parameter. One model enters in the quintessence model 
category (DE5), the second one is a phantom model (DE6) and the last one (DE7) is characterised by the barrier 
crossing, i.e., the model considered shows a phantom regime at low redshifts ($w<-1$) and a quintessence regime at 
earlier times ($w>-1$).

As previously stated, quintessence models are described by a scalar field not interacting with matter and are fully 
described by a kinetic and a potential energy term. Since the nature of dark energy is unknown, the potential has an 
ad hoc functional form and its second derivative represents the mass of the scalar field. These models naturally have 
an evolving dark energy equation of state $w$. Scalar fields are therefore viable candidates for the dark energy 
component.

Phantom models instead have $w<-1$ and challenge the foundations of theoretical physics violating several energy 
conditions. Phantom models have a negative defined kinetic energy term and due to the super-negative equation of 
state, the energy budget of the Universe gets completely dominated by them in the future.

The solution of Equation~\ref{eqn:deceq} is (for a generic time-dependant equation-of-state parameter $w(a)$)
\begin{equation}\label{eqn:rhoa}
 \rho(a)=\rho(a=1) e^{-3\int_1^a [1+w(a^\prime)]d\ln a^\prime}\;.
\end{equation}
In the particular case of constant equation of state, Equation~\ref{eqn:rhoa} reduces to
\begin{equation}\label{eqn:rho_const}
 \rho(a)=\rho(a=1)a^{-3(1+w)}\;,
\end{equation}
where it appears clearly that for the cosmological constant $\rho(a)=\rho(a=1)$, hence the name.

In this work we will use the following cosmological parameters (recall that the curvature is null): 
$\Omega_{\rm m}=0.32$ and $\Omega_{\rm de}=0.68$, while $h=0.72$, in agreement with recent determinations by Planck 
\citep{Planck2013_XVI} for flat $\Lambda$CDM models. The normalization of the power spectrum for the $\Lambda$CDM 
model is $\sigma_8=0.776$.

\begin{figure}
 \centering
 \includegraphics[scale=0.5,angle=-90]{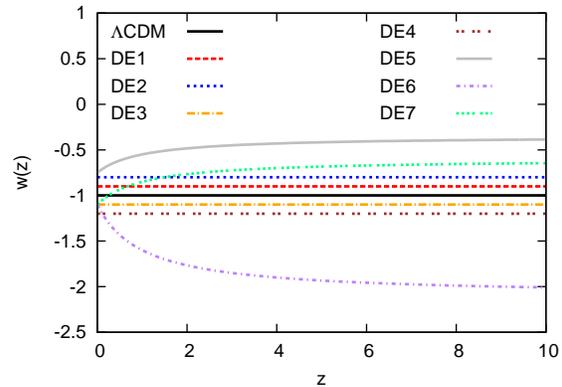}
 \caption[Equation of state]{\small Time evolution of the equation-of-state parameter as a function of the redshift 
$z$ for the dark energy models studied in this work. The black, red, blue, orange and brown lines represent the 
$\Lambda$CDM, the DE1, the DE2, the DE3 and the DE4 model, respectively. The grey line represents the DE5 model, the 
purple line the DE6 model and the green line the DE7 model.}
 \label{fig:wz}
\end{figure}

In table~\ref{tab:wparam} we give the values of the parameters describing the equations of state of the models 
considered here both for a null time evolution (models DE$n$, with $n$ from 1 to 4) and for a time evolution (models 
DE$n$, with $n$ from 5 to 7) of the equation of state. We recall that for $w_a=0$, the CPL parametrization reduces to 
a constant equation of state with $w=w_0$. We show the time evolution of the equation-of-state parameter $w(z)$ in 
Fig.~\ref{fig:wz}.

\begin{table}
 \centering
 \caption[Equations of state]{Values of the parameters describing the equations of state considered in this work.}
 \begin{tabular}{lcc}
  \hline
  \hline
  Model & $w_0$ & $w_a$\\
  \hline
  $\Lambda$CDM & -1 & 0 \\
  DE1 & -0.9 & 0 \\
  DE2 & -0.8 & 0 \\
  DE3 & -1.1 & 0 \\
  DE4 & -1.2 & 0 \\
  DE5 & -0.75 & 0.4 \\
  DE6 & -1.1 & -1. \\
  DE7 & -1.1 & 0.5 \\
  \hline
 \label{tab:wparam}
 \end{tabular}
 \begin{flushleft}
  \vspace{-0.5cm}
  {\small}
 \end{flushleft}
\end{table}

For the models DE$n$, with $n=5,6,7$, the variation of the equation of state as a function of the redshift is quite 
mild, with major variations for $z\lesssim 2$. At higher redshifts all the three models reach a constant value for the 
equation of state, being $w=-0.35,-2,-0.6$ for the models DE5, DE6, DE7, respectively. We also point out that the 
barrier crossing for the model DE7 takes place for $z\approx 0.25$, having a phantom (quintessence) behaviour for 
smaller (higher) redshifts. We checked that all the models do not have an appreciable amount of dark energy at early 
times, therefore they can not be considered as belonging to the class of early dark energy models.

Perturbations for dark energy are described by the effective sound speed $c^2_{\rm eff}$ that relates density 
perturbations to pressure perturbations via the relation $\delta p=c^2_{\rm eff}\delta\rho c^2$. In the following we 
will consider two different values for the effective sound speed, usually assumed in literature: $c^2_{\rm eff}=0$ 
(clustering DE) and $c^2_{\rm eff}=1$ (smooth DE). Canonical scalar fields have $c^2_{\rm eff}=1$, whereas models with 
vanishing $c^2_{\rm eff}$ can be build from k-essence models \citep{Armendariz-Picon2001,Chimento2005,Creminelli2009} 
or two scalar fields \citep{Lim2010}. We will show in Sect.~\ref{sect:escm} how this term enters into the equations 
for the (extended) spherical collapse model.

\section{Extended Spherical Collapse Model (ESCM)}\label{sect:escm}
In this section we review the basic formalism used to derive the equations for the spherical collapse model and how 
this can be extended to include shear and rotation terms.

The basic assumption in the framework of the spherical collapse model is that objects form under the gravitational 
collapse of spherical dark matter overdense perturbations. This is clearly a rather crude assumption because it is 
known that primordial seeds are not spherical, but they are triaxial and rotate 
\citep[see e.g.][]{Bardeen1986,DelPopolo2001,DelPopolo2002,Shaw2006,Bett2007}. Nevertheless the model accurately 
reproduces the results of N-body simulations.

The spherical and ellipsoidal collapse models were extensively investigated in literature 
\citep[see, e.g.][]{Bernardeau1994,Bardeen1986,Ohta2003,Ohta2004,Basilakos2009,Pace2010,Basilakos2010} assuming that 
dark energy perturbations are negligible, while other studies took into account also the effects of perturbations for 
the dark energy fluid 
\citep[see][]{Mota2004,Nunes2006,Abramo2007,Abramo2008,Abramo2009a,Abramo2009b,Creminelli2010,Basse2011,Batista2013}. 
More recently, the spherical collapse model was extended to investigate coupled 
\citep{Pettorino2008,Wintergerst2010a,Tarrant2012} and extended dark energy (scalar-tensor) models 
\citep{Pettorino2008,Pace2014}.

While the general equations including the shear and rotation were explicitly written in the case of smooth dark 
energy \citep{Pace2010} and for clustering dark energy \citep{Abramo2007}, the effects of these two nonlinear terms 
were investigate only recently in \cite{DelPopolo2013a,DelPopolo2013b} for smooth dark energy models and in 
\cite{DelPopolo2013c} for Chaplygin cosmologies.

Following \cite{Abramo2007} and \cite{Abramo2008}, the full perturbed equations describing the evolution of the dark 
matter ($\delta_{\rm DM}$) and dark energy ($\delta_{\rm DE}$) perturbations are:

\begin{eqnarray}
 \delta_{\rm DM}^{\prime}+\left(1+\delta_{\rm DM}\right)\frac{\theta_{\rm DM}}{aH} & = & 0\;,\label{eqn:DMp}\\
 \delta_{\rm DE}^{\prime}-\frac{3}{a}w\delta_{\rm DE}+
 \left[1+w+\delta_{\rm DE}\right]\frac{\theta_{\rm DE}}{aH} & = & 0\;.\label{eqn:DEp}
\end{eqnarray}
In the previous equations, $w$ represents the equation of state of dark energy at the background level and the prime 
is the derivative with respect to the scale factor. The two variables $\theta_{\rm DM}$ and $\theta_{\rm DE}$ are the 
divergence of the peculiar velocity for the dark matter and the dark energy component, respectively. 
Equation~\ref{eqn:DEp} is valid in the limit $c^{2}_{\rm eff}=0$, the limit of clustering dark energy. For the case 
$c_{\rm eff}^2>0$, dark energy perturbations are usually negligible on small scales, as shown for example in 
\cite{Batista2013}.

To determine the equation for the evolution of the divergence of the peculiar velocity we have to make some 
assumptions on the influence of the shear and rotation terms on the perturbations of the two fluids considered. If we 
assume that only dark matter experiences shear and rotation terms, then the two peculiar velocities are different 
($\theta_{\rm DM}\neq\theta_{\rm DE}$) and we will have two different equations: one including shear and rotation for 
the dark matter and one for the dark energy component without the extra terms. If instead we assume that DE 
experiences the effects of the shear and the rotation terms in the same fashion of the DM component, then the two 
peculiar velocities will be the same and we need to solve a single differential equation.\\
Here we explicitly write the two different equations for the peculiar velocities and in the next sections we will 
study the consequences of this assumption. Having therefore two different Euler equations, the equations for the 
divergence of the peculiar velocities are:

\begin{eqnarray}
 \theta_{\rm DM}^{\prime}+\frac{2}{a}\theta_{\rm DM}+\frac{\theta_{\rm DM}^2}{3aH}+
 \frac{\sigma_{\rm DM}^2-\omega_{\rm DM}^2}{aH}+&\nonumber\\
 \frac{3H}{2a}\left[\Omega_{\rm DM}\delta_{\rm DM}+\Omega_{\rm DE}\delta_{\rm DE}\right]&=&0\;,\label{eqn:thetaDMp}\\
 \theta_{\rm DE}^{\prime}+\frac{2}{a}\theta_{\rm DE}+\frac{\theta_{\rm DE}^2}{3aH}+
 \frac{3H}{2a}\left[\Omega_{\rm DM}\delta_{\rm DM}+\Omega_{\rm DE}\delta_{\rm DE}\right]&=&0\;.\label{eqn:thetaDEp}
\end{eqnarray}

The shear tensor $\sigma_{ij}$ and the vorticity tensor $\omega_{ij}$ are defined as
\begin{eqnarray}
 \sigma_{ij} & = & \frac{1}{2}\left(\frac{\partial u^j}{\partial x^i}+\frac{\partial u^i}{\partial x^j}\right)
 -\frac{1}{3}\theta\delta_{ij}\;, \\
 \omega_{ij} & = & \frac{1}{2}\left(\frac{\partial u^j}{\partial x^i}-\frac{\partial u^i}{\partial x^j}\right)\;.
\end{eqnarray}
The terms $\sigma^2$ and $\omega^2$ represent the contractions of the tensors $\sigma_{ij}$ and $\omega_{ij}$, 
respectively.

It is convenient to consider a dimensionless divergence of the comoving peculiar velocity, defined as 
$\tilde{\theta}=\theta/H$. Therefore Eqs.~\ref{eqn:DMp}-\ref{eqn:thetaDEp} read now

\begin{eqnarray}
 \delta_{\rm DM}^{\prime}+\left(1+\delta_{\rm DM}\right)\frac{\tilde{\theta}_{\rm DM}}{a} & = & 0\;,\label{eqn:DMpt}\\
 \delta_{\rm DE}^{\prime}-\frac{3}{a}w\delta_{\rm DE}+
 \left[1+w+\delta_{\rm DE}\right]\frac{\tilde{\theta}_{\rm DE}}{a} & = & 0\;,\label{eqn:DEpt}\\
 \tilde{\theta}_{\rm DM}^{\prime}+\left(\frac{2}{a}+\frac{H^{\prime}}{H}\right)\tilde{\theta}_{\rm DM}+
 \frac{\tilde{\theta}_{\rm DM}^2}{3a}+&\nonumber\\
 \frac{\tilde{\sigma}_{\rm DM}^2-\tilde{\omega}_{\rm DM}^2}{a}+
 \frac{3}{2a}\left[\Omega_{\rm DM}\delta_{\rm DM}+\Omega_{\rm DE}\delta_{\rm DE}\right]&=&0\;,\label{eqn:thetaDMpt}\\
 \tilde{\theta}_{\rm DE}^{\prime}+\left(\frac{2}{a}+\frac{H^{\prime}}{H}\right)\tilde{\theta}_{\rm DE}+
 \frac{\tilde{\theta}_{\rm DE}^2}{3a}+&\nonumber\\
 \frac{3}{2a}\left[\Omega_{\rm DM}\delta_{\rm DM}+\Omega_{\rm DE}\delta_{\rm DE}\right]&=&0\;.\label{eqn:thetaDEpt}
\end{eqnarray}
We remind the reader that this set of equations is valid when dark energy is not affected by shear and rotation, 
otherwise Eqs.~\ref{eqn:thetaDMpt} and~\ref{eqn:thetaDEpt} will be identical and $\theta_{\rm DM}=\theta_{\rm DE}$.

To solve the system of equations \ref{eqn:DMpt}, \ref{eqn:DEpt}, \ref{eqn:thetaDMpt} and \ref{eqn:thetaDEpt}, it is 
necessary to determine the initial conditions. At early times, the aforementioned system of equations can be 
linearised and it reads
\begin{eqnarray}
 \delta_{\rm DM}^{\prime} & = & -\frac{\tilde{\theta}_{\rm DM}}{a}\;,\label{eqn:DMptl}\\
 \delta_{\rm DE}^{\prime}-\frac{3}{a}w\delta_{\rm DE}& = & -(1+w)\frac{\tilde{\theta}_{\rm DE}}{a}\;,\label{eqn:DEptl}\\
 \tilde{\theta}_{\rm DM}^{\prime}+\left(\frac{2}{a}+\frac{H^{\prime}}{H}\right)\tilde{\theta}_{\rm DM} & = &
 -\frac{3}{2a}\left[\Omega_{\rm DM}\delta_{\rm DM}+\Omega_{\rm DE}\delta_{\rm DE}\right]\;,\label{eqn:thetaDMptl}\\
 \tilde{\theta}_{\rm DE}^{\prime}+\left(\frac{2}{a}+\frac{H^{\prime}}{H}\right)\tilde{\theta}_{\rm DE} & = &
 -\frac{3}{2a}\left[\Omega_{\rm DM}\delta_{\rm DM}+\Omega_{\rm DE}\delta_{\rm DE}\right]\;.\label{eqn:thetaDEptl}
\end{eqnarray}
Hence at the linear level, the peculiar velocity perturbations are identical for both fluids.

The initial value for the dark matter overdensity can be found as outlined in \cite{Pace2010,Pace2012,Pace2014} and 
\cite{Batista2013}. Here we just recall the general procedure. Since at collapse time $a_{\rm c}$ 
the collapsing sphere reduces to a point, its density is formally infinite. Therefore the initial overdensity 
$\delta_{\rm DM,i}$ is given by the value such that $\delta_{\rm DM}\rightarrow +\infty$ for 
$a\rightarrow a_{\rm c}$. Knowing $\delta_{\rm DM,i}$ and assuming that at early times it behaves as a power law, 
$\delta_{\rm DM}=Aa^n$, it is easy to evaluate the initial amplitude for the dark energy and the peculiar velocity 
perturbations:
\begin{eqnarray}
 \delta_{\rm DE,i} & = & \frac{n}{(n-3w)}(1+w)\delta_{\rm DM,i}\;,\label{eqn:deltadei}\\
 \tilde{\theta}_{\rm DM,i} & = & -n\delta_{\rm DM,i}\;.\label{eqn:thetaDMi}\\
 \tilde{\theta}_{\rm DE,i} & = & \tilde{\theta}_{\rm DM,i}\;.\label{eqn:thetaDEi}
\end{eqnarray}
For an EdS model, $n=1$, but in general deviations for DE models are very small, even for early dark energy models 
\citep{Ferreira1998,Batista2013}.

To evaluate the functional form of the term $\sigma_{\rm DM}^2-\omega_{\rm DM}^2$ we refer to the works by 
\cite{DelPopolo2013a,DelPopolo2013b} and we define the quantity $\alpha$ as the ratio between the rotational and 
the gravitational term
\begin{equation}
 \alpha=\frac{L^2}{M^3RG}\;,
\end{equation}
where $M$ and $R$ are the mass and the radius of the spherical overdensity respectively and $L$ its angular momentum. 
Values for $\alpha$ range from 0.05 for galactic masses ($M\approx 10^{11}~M_{\odot}/h$) to $3\times 10^{-6}$ for 
cluster scales ($M\approx 10^{15}~M_{\odot}/h$).

As explained in \cite{DelPopolo2013b} the basic assumption here made is that the collapse preserves the value of the 
ratio of the acceleration between the shear rotation term and the gravitational field. This is a reasonable assumption 
as explained in \cite{DelPopolo2013b}. As shown in \cite{DelPopolo2013c}, based on the above outlined argument for the 
definition of the rotation term, the additional term in the equations for the spherical collapse model (see 
Equation~\ref{eqn:thetaDMpt}) is
\begin{equation}
 \tilde{\sigma}_{\rm DM}^2-\tilde{\omega}_{\rm DM}^2=
 -\frac{3}{2}\alpha\Omega_{\rm DM}\delta_{\rm DM}\;.
\end{equation}
According to this Ansatz, Equation~\ref{eqn:thetaDMpt} now reads
\begin{eqnarray}\label{eqn:thetaDMtp_alpha}
 \tilde{\theta}_{\rm DM}^{\prime}+\left(\frac{2}{a}+\frac{H^{\prime}}{H}\right)\tilde{\theta}_{\rm DM} 
+\frac{\tilde{\theta}_{\rm DM}^2}{3a}+\nonumber\\
 \frac{3}{2a}\left[(1-\alpha)\Omega_{\rm DM}\delta_{\rm DM}+\Omega_{\rm DE}\delta_{\rm DE}\right]=0\;.
\end{eqnarray}
If instead also dark energy is affected by shear and rotation, in the same way as dark matter, both velocity fields 
are determined by the equation:
\begin{equation}\label{eqn:thetatp_alpha}
 \tilde{\theta}^{\prime}+\left(\frac{2}{a}+\frac{H^{\prime}}{H}\right)\tilde{\theta}+\frac{\tilde{\theta}^2}{3a}+
 \frac{3}{2a}(1-\alpha)\left[\Omega_{\rm DM}\delta_{\rm DM}+\Omega_{\rm DE}\delta_{\rm DE}\right]=0\;.
\end{equation}

\section{Results}\label{sect:results}
In this section we present results for the linear and nonlinear evolution of perturbations. We first start with 
quantities derived in the framework of the spherical collapse model and we continue with a discussion of how rotation 
and shear affect the mass function in clustering dark energy cosmologies.\\
As shown in \cite{Batista2013}, the main difficulty is to study the evolution of dark energy perturbations in the 
non-linear regime \citep[see also][]{Mota2004,Nunes2006,Abramo2007,Creminelli2010,Basse2011}.\\
\cite{Batista2013} clearly demonstrated that dark energy fluctuations are very sensitive to the value of 
$c^{2}_{\rm eff}$: when $c^{2}_{\rm eff}=1$, on small scales, where non-linear evolution is important, dark energy 
fluctuations are negligible with respect to the dark matter fluctuations $\delta_{\rm DM}$ therefore ignoring them 
when solving the system of equations describing the ESCM (Equations~\ref{eqn:DMpt},~\ref{eqn:DEpt},
~\ref{eqn:thetaDEpt} and~\ref{eqn:thetaDMtp_alpha}) does not introduce any significant error. Different is the 
situation when $c^{2}_{\rm eff}=0$ since DE fluctuations can be comparable to the DM ones. In this case we cannot 
neglect them, otherwise the error introduced will be significant and invalidate our results and conclusions.

\subsection{Parameters of the ESCM}\label{sect:escm_params}
The two main quantities that can be evaluated working within the framework of the ESCM are the linear overdensity 
parameter $\delta_{\rm c}$ and the virial overdensity $\Delta_{\rm V}$. The linear overdensity parameter is a 
fundamental theoretical quantity entering, together with the linear growth factor, into analytical formulations of the 
mass function \citep[see e.g.][]{Press1974,Sheth2001,Sheth2002}. The virial overdensity instead, is used to define the 
size of halos when considered spherical. Given a halo of mass $M$, it represents the mean density enclosed in the 
radius $R$ and the mass and the radius are related to each other via the relation 
$M=4/3\pi\bar{\rho}(z)\Delta_{\rm V}(z) R^3$ where $\bar{\rho}(z)=\bar{\rho}_{,0}(1+z)^3$ is the mean matter density 
in the Universe.

\begin{figure}
 \centering
 \includegraphics[scale=0.45,angle=-90]{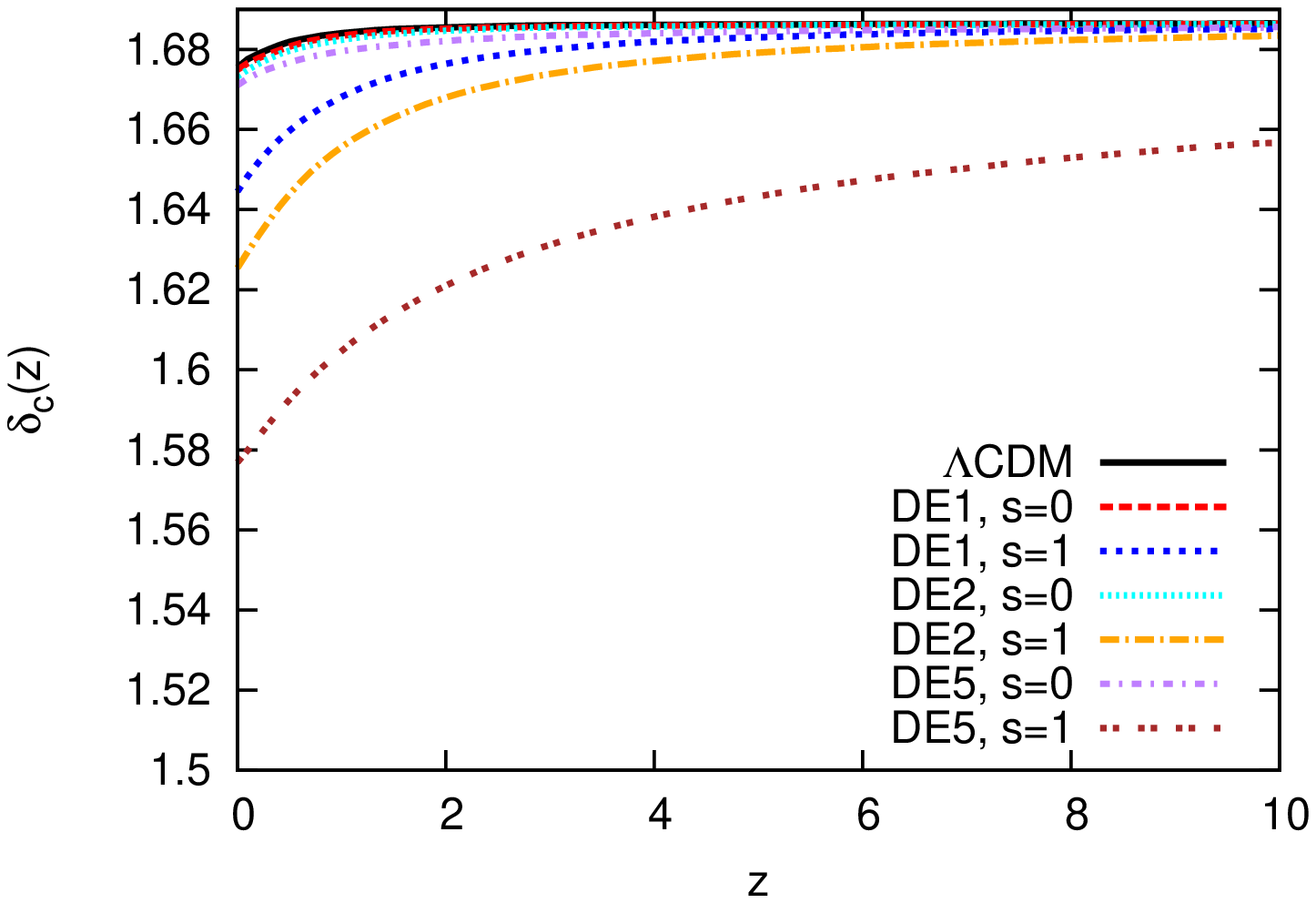}
 \includegraphics[scale=0.45,angle=-90]{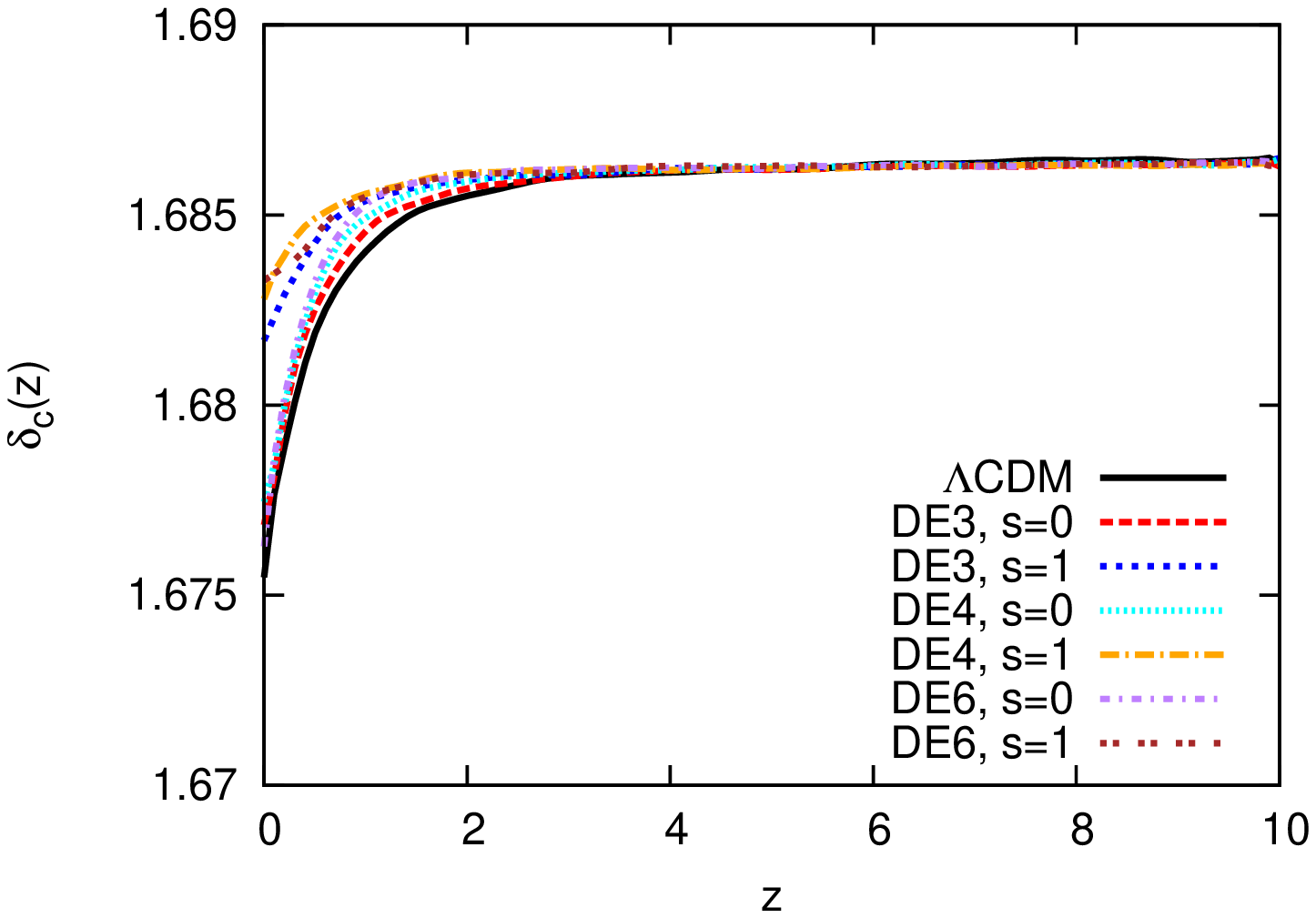}
 \includegraphics[scale=0.45,angle=-90]{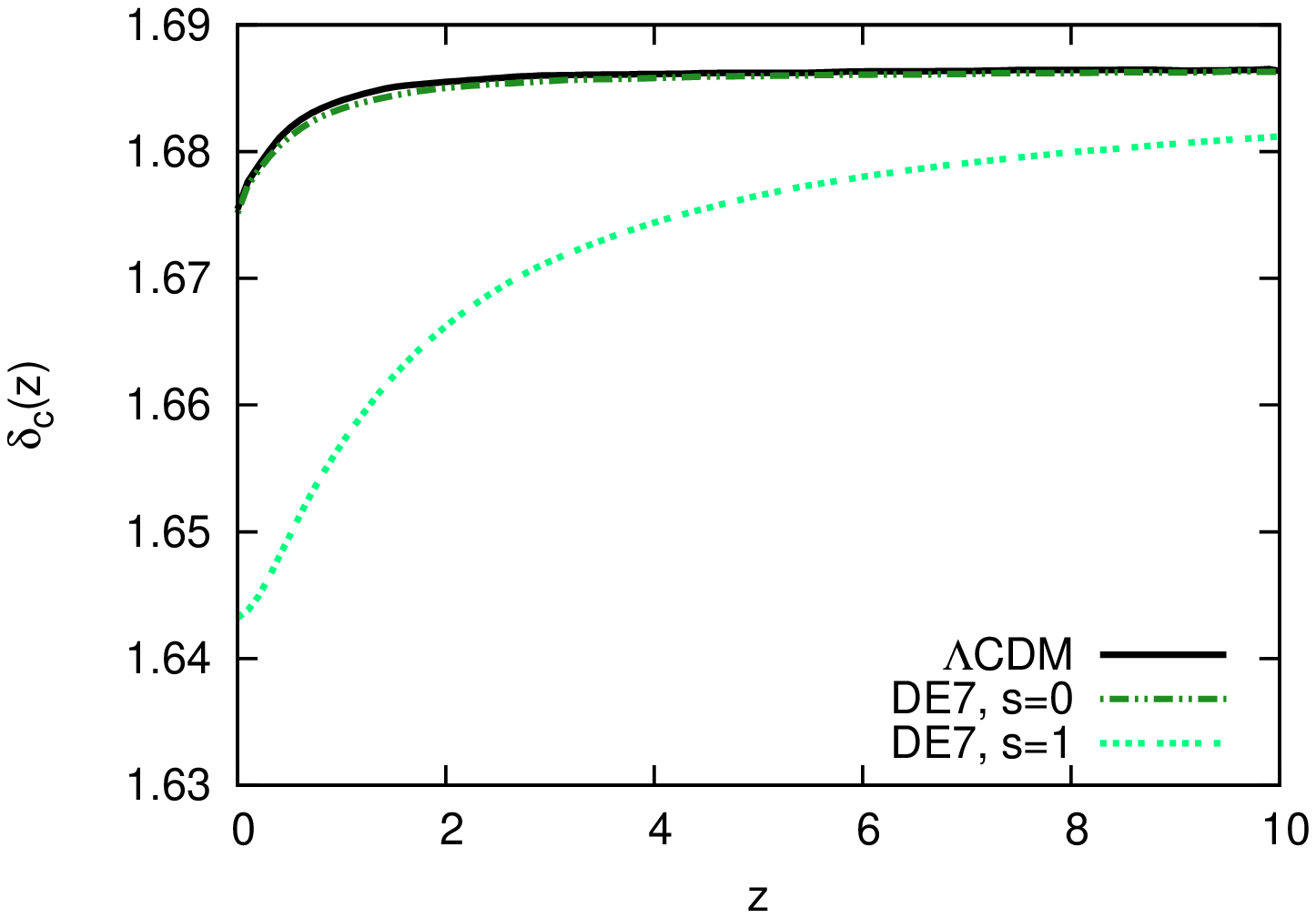}
 \caption{Linear overdensity parameter $\delta_{\rm c}(z)$ for quintessence (upper panel), phantom (middle panel) and 
models with barrier crossing (bottom panel). In the upper panel, model DE1 with 
$c^{2}_{\rm eff}=0$ ($c^{2}_{\rm eff}=1$) is shown with a dashed red (blue short dashed) curve, model DE2 with 
$c^{2}_{\rm eff}=0$ ($c^{2}_{\rm eff}=1$) is shown with cyan dotted (yellow dot-dashed) curve, model DE5 with 
$c^{2}_{\rm eff}=0$ ($c^{2}_{\rm eff}=1$) with a violet dotted short-dashed (brown dot-dotted) curve. In the middle 
panel, model DE3 with $c^{2}_{\rm eff}=0$ ($c^{2}_{\rm eff}=1$) is shown with a red dashed (blue short-dashed) curve, 
model DE4 with $c^{2}_{\rm eff}=0$ ($c^{2}_{\rm eff}=1$) is shown with a cyan dotted (yellow dot-dashed) curve while 
model DE6 with $c^{2}_{\rm eff}=0$ ($c^{2}_{\rm eff}=1$) is shown with violet dotted short-dashed (brown dot-dotted) 
curve. In the bottom panel, model DE7 with $c^{2}_{\rm eff}=0$ ($c^{2}_{\rm eff}=1$) is shown with a dark green 
dot-dotted short-dashed (light green dot-dot-dotted) curve. In all the panels, a solid black line shows the 
$\Lambda$CDM model. For simplicity, we used the notation $s=c^{2}_{\rm eff}$ in the labels.}
 \label{fig:deltac_norot}
\end{figure}

Once the initial conditions are found, we can evolve Equations~\ref{eqn:DMptl}-\ref{eqn:thetaDEptl} from the initial 
time $a_{\rm i}\approx 10^{-5}$ to the collapse time $a_{\rm c}$. This function therefore depends on both the linear 
and non-linear evolution of perturbations.

In Figure~\ref{fig:deltac_norot} we show the time evolution of the linear overdensity parameter $\delta_{\rm c}$ for 
the usual case when shear and rotations are not included. We do so in order to better show how the additional terms 
modify this function. We show our results grouping the models as quintessence (top panel), phantom (middle panel) and 
barrier crossing models (bottom panel). We refer to the caption for line styles and colours of each model.

\begin{figure}
 \centering
 \includegraphics[scale=0.45,angle=-90]{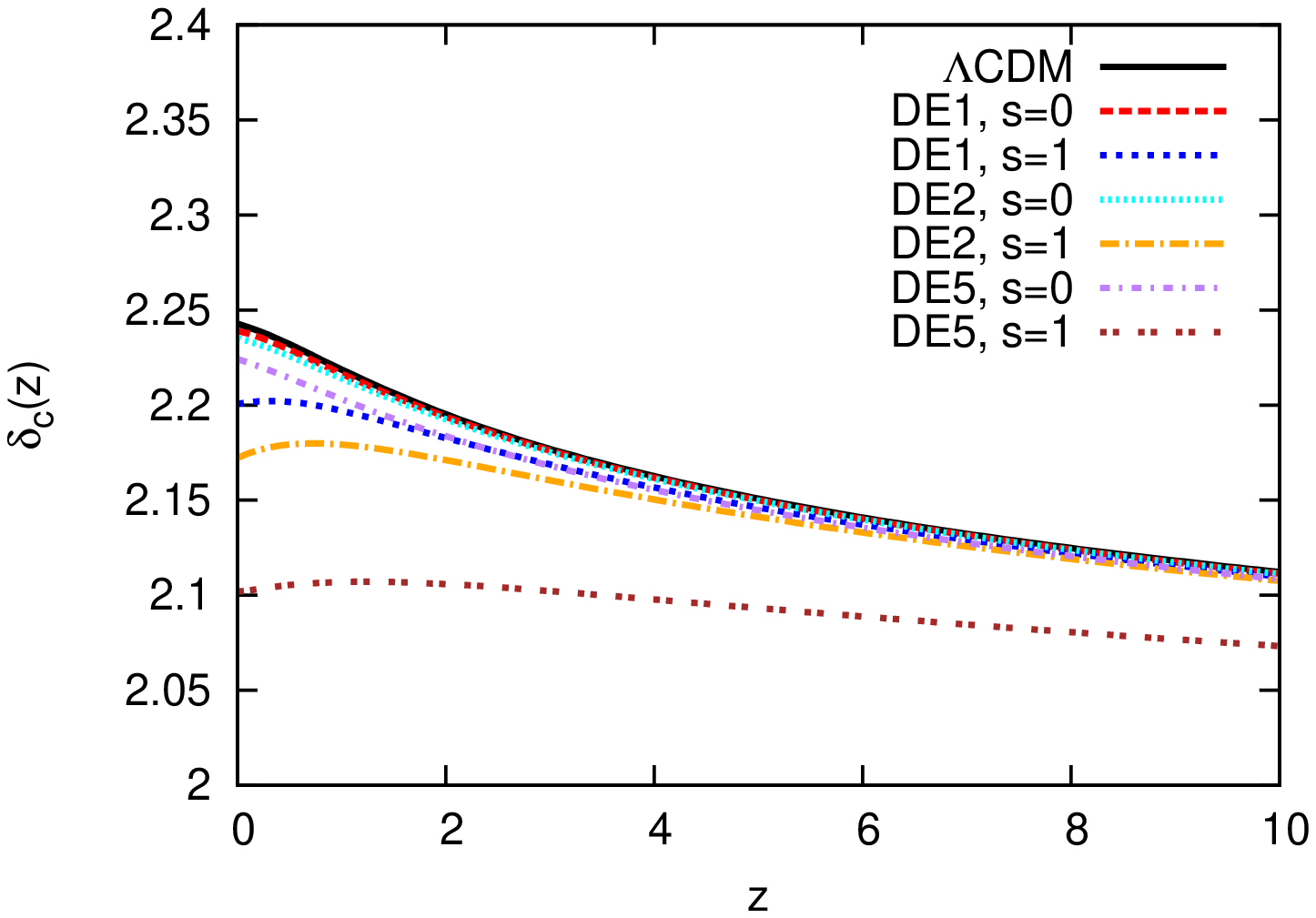}
 \includegraphics[scale=0.45,angle=-90]{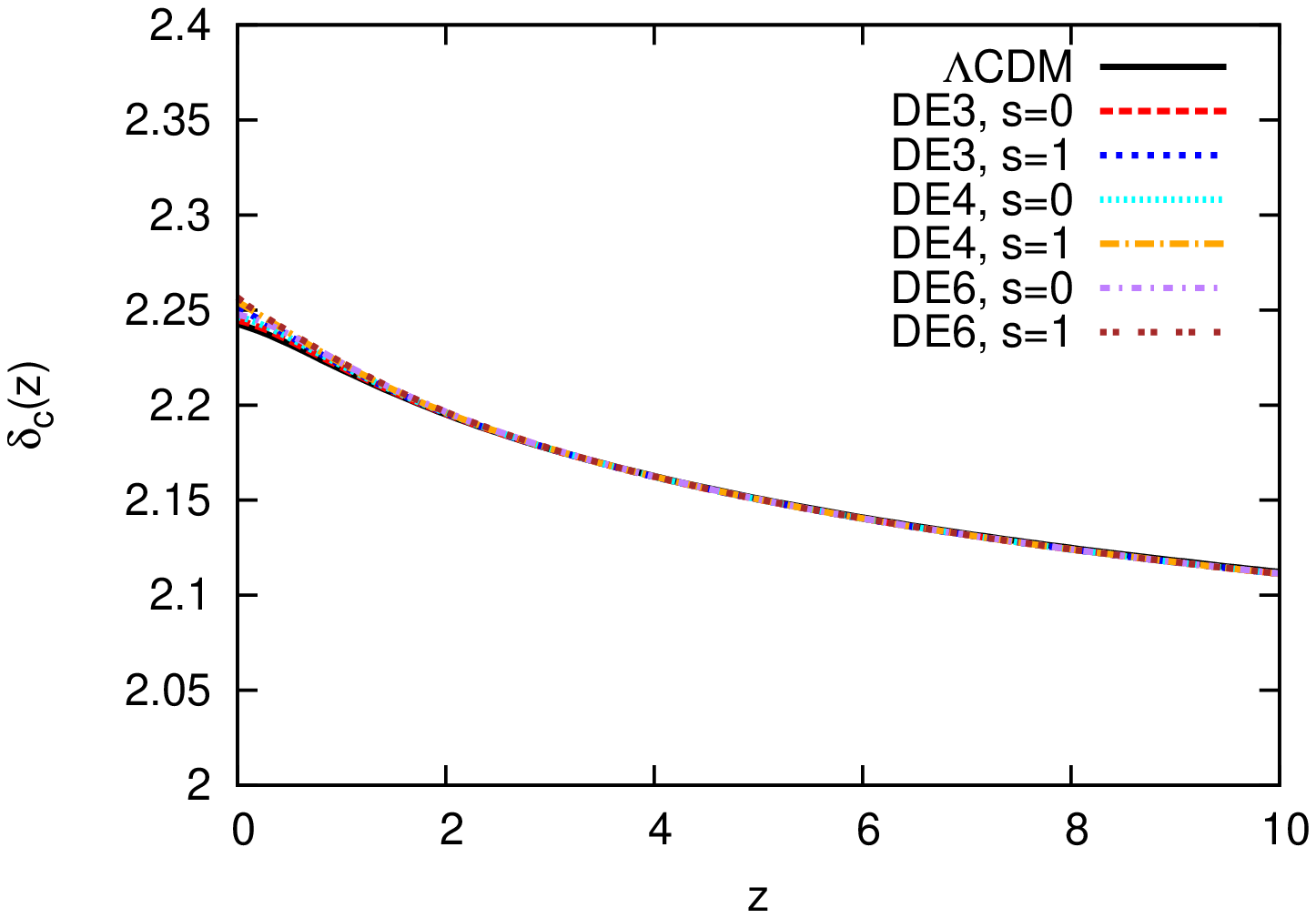}
 \includegraphics[scale=0.45,angle=-90]{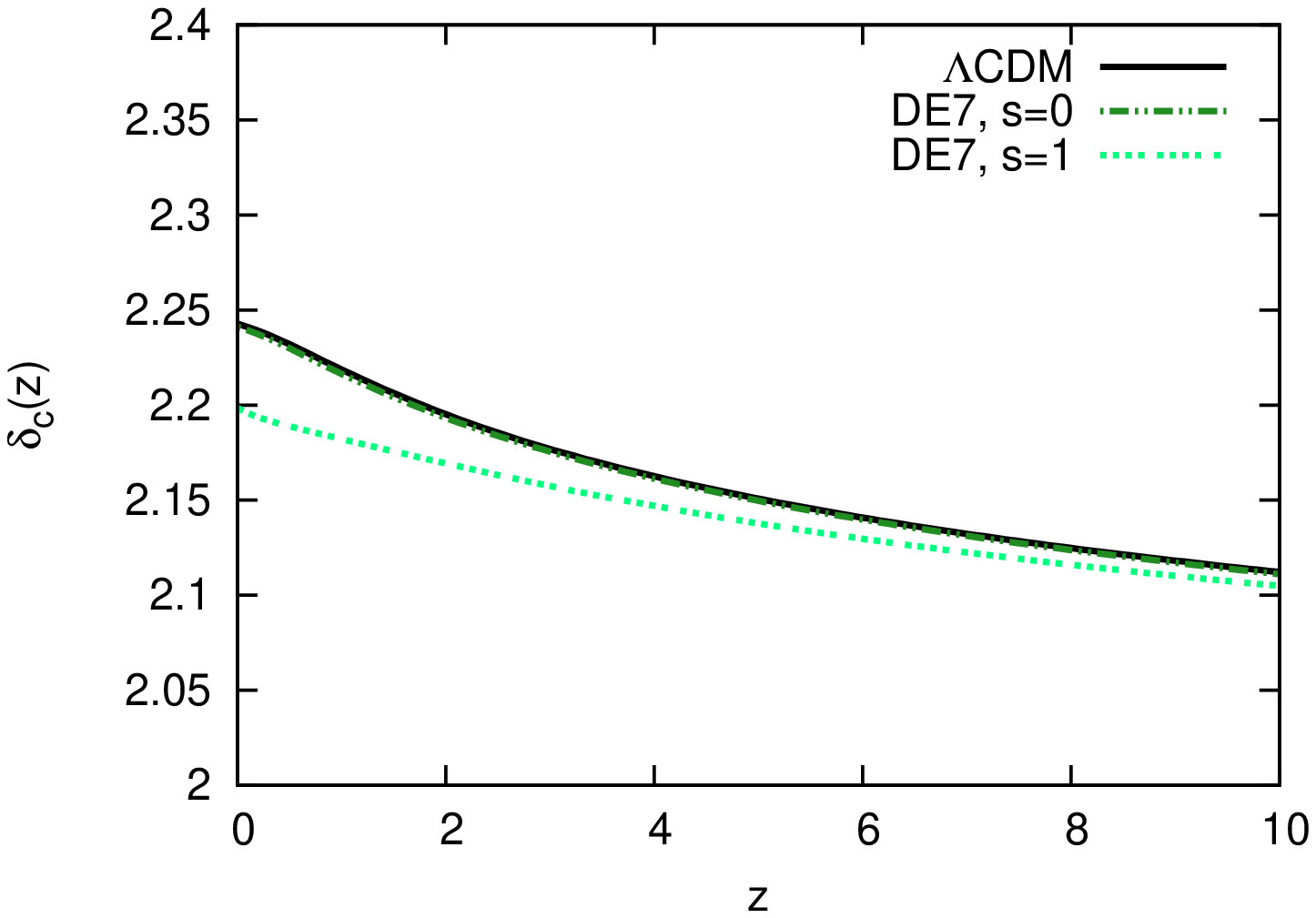}
 \caption{Linear overdensity parameter $\delta_{\rm c}(z)$ for quintessence (upper panel), phantom (middle panel) and 
models with barrier crossing (bottom panel) including the shear and rotation terms in the equations for the evolution 
of the perturbations. Panels refer to galactic scale mass objects ($M\approx 10^{11}~M_{\odot}/h$), corresponding to 
$\alpha=0.05$. Line styles and colours are the same as in Fig.~\ref{fig:deltac_norot}.}
 \label{fig:deltac_rot}
\end{figure}

\begin{figure}
 \centering
 \includegraphics[scale=0.45,angle=-90]{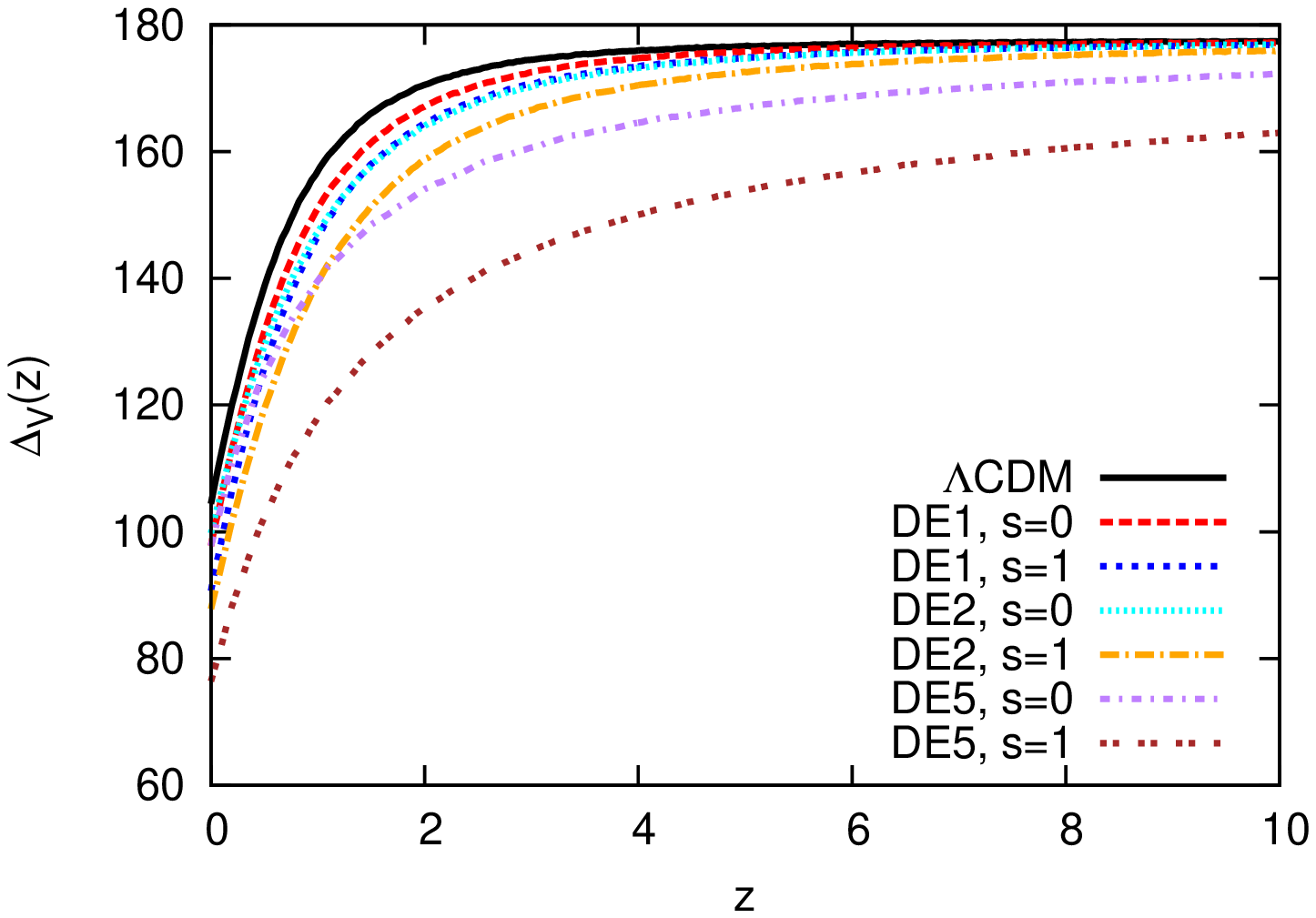}
 \includegraphics[scale=0.45,angle=-90]{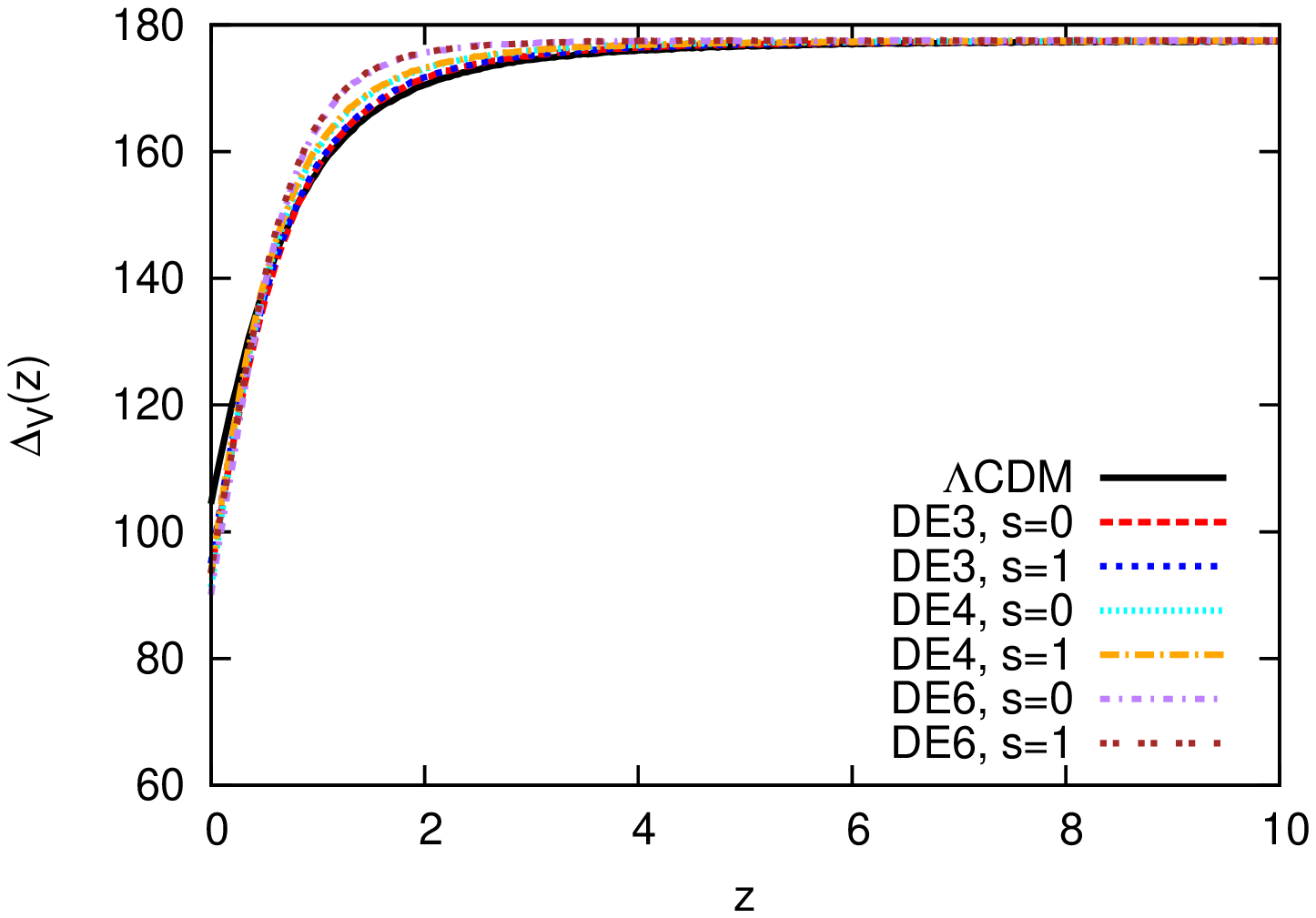}
 \includegraphics[scale=0.45,angle=-90]{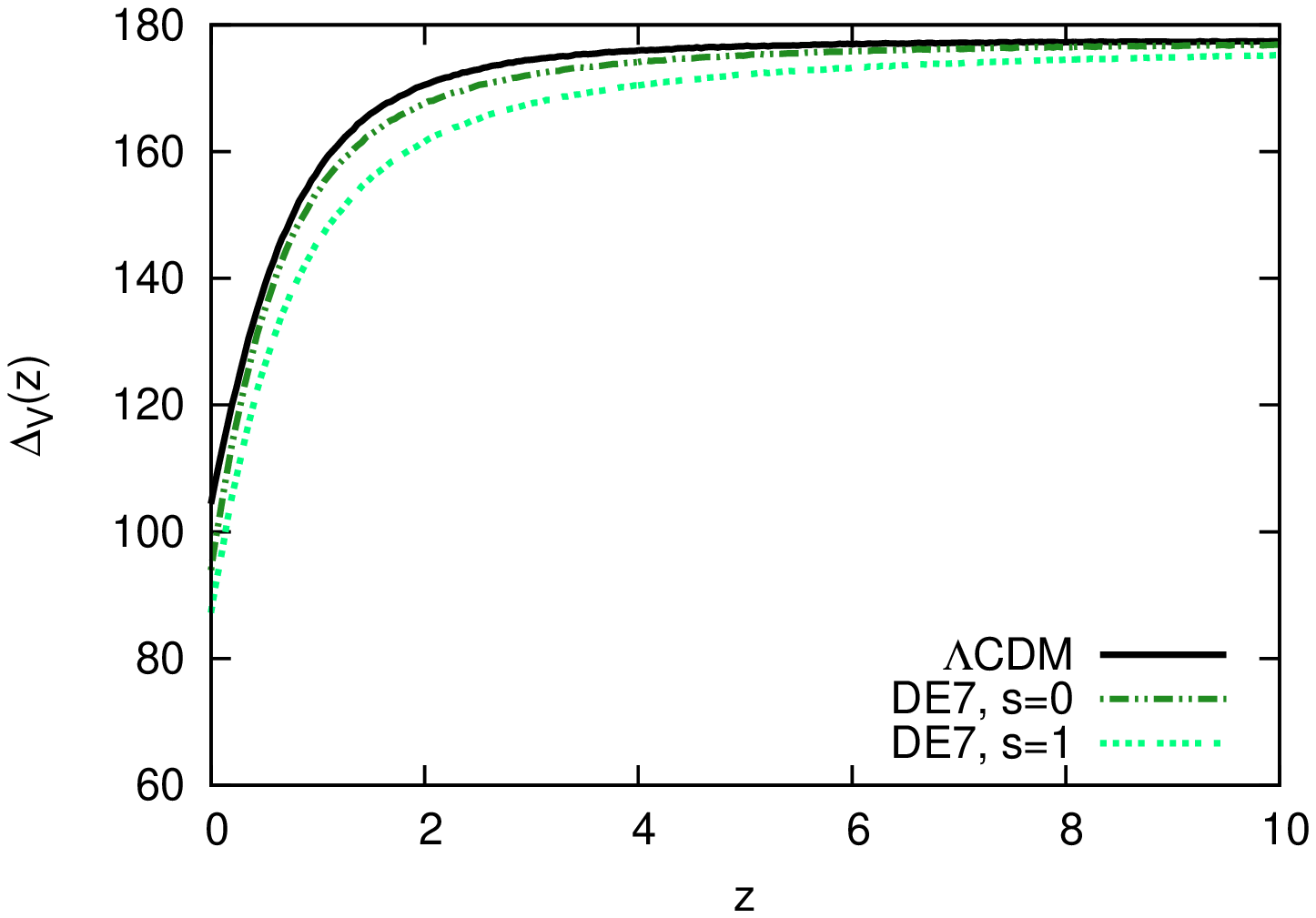}
 \caption{Virial overdensity for the difference DE models. In the upper (middle) panel we show results for the 
quintessence (phantom) models while in the bottom panel we present results for the model with barrier crossing. Line 
styles and colours are as in Figure~\ref{fig:deltac_norot}.}
 \label{fig:deltav_norot}
\end{figure}

\begin{figure}
 \centering
 \includegraphics[scale=0.45,angle=-90]{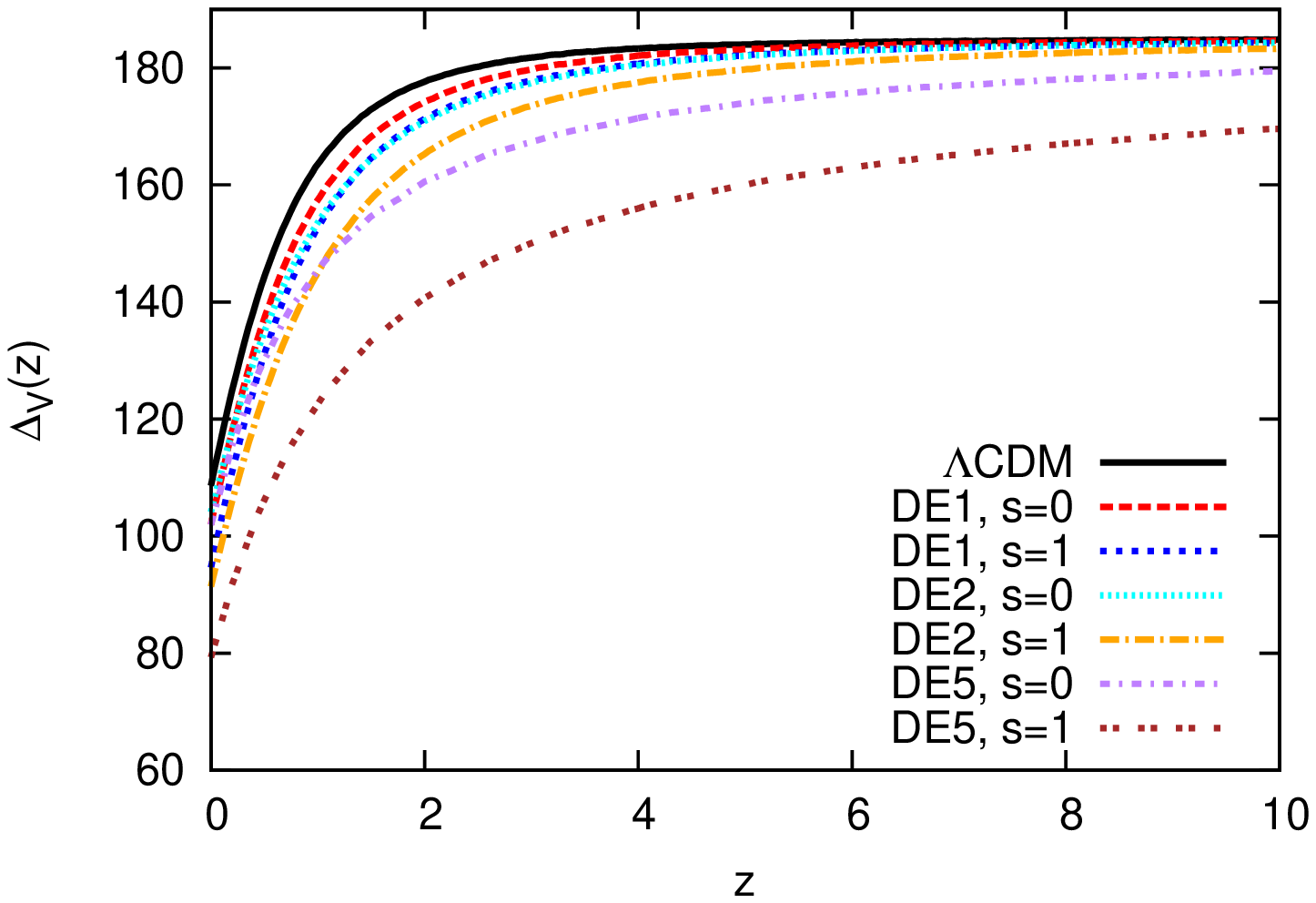}
 \includegraphics[scale=0.45,angle=-90]{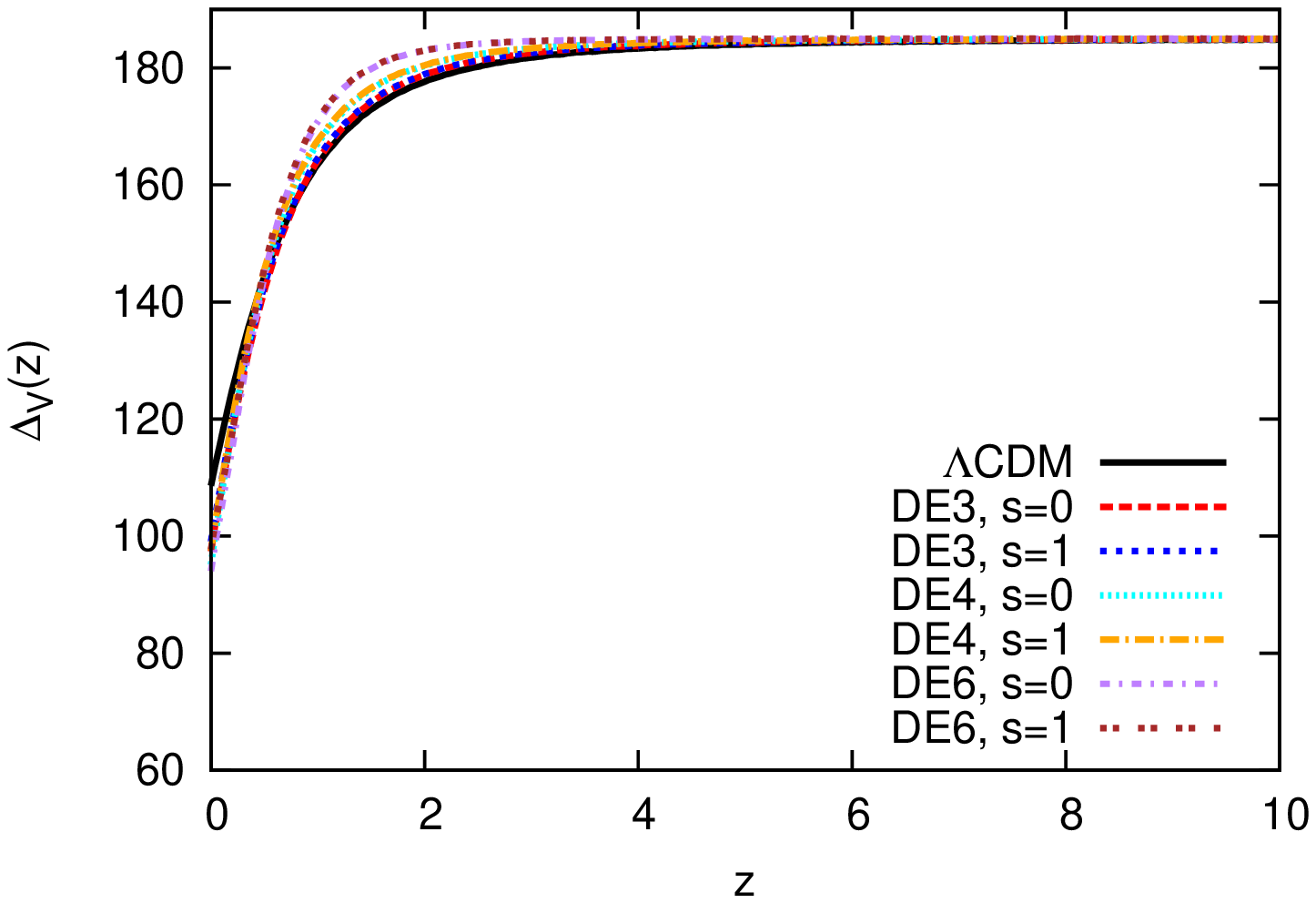}
 \includegraphics[scale=0.45,angle=-90]{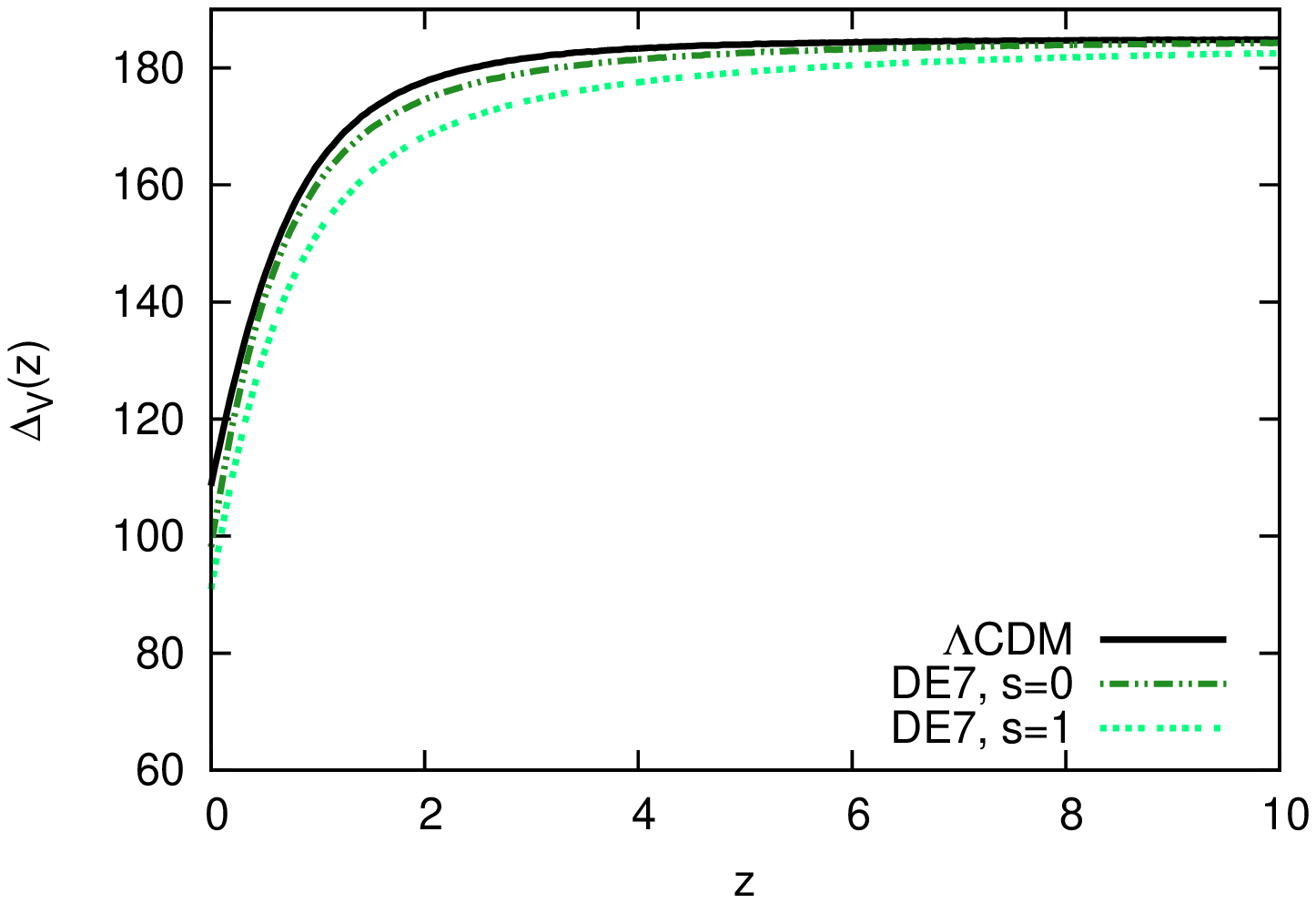}
 \caption{Virial overdensity for the difference DE models with shear and rotation terms included. In the upper 
 (middle) panel we show results for the quintessence (phantom) models while in the bottom panel we present results 
 for the model with barrier crossing. Panels refer to galactic scale mass objects ($M\approx 10^{11}~M_{\odot}/h$), 
corresponding to $\alpha=0.05$. Line styles and colours are as in Figure~\ref{fig:deltac_norot}.}
 \label{fig:deltav_rot}
\end{figure}

The first important point to highlight is that quintessence models ($w\geqslant -1$) always show a lower 
$\delta_{\rm c}(z)$ with respect to the $\Lambda$CDM model, while the phantom models always have a higher value, due 
to the fastest accelerated expansion of the universe, that obstacles structure formation. We also notice that models 
with $c^{2}_{\rm eff}=0$ are more similar to the $\Lambda$CDM model than for the case with $c^{2}_{\rm eff}=1$, in 
agreement with what was found by \cite{Batista2013} for early dark energy models, where we refer for a deeper 
explanation. For our purposes, it suffices to recall that this happens because DE perturbations contribute to the 
gravitational potential via the Poisson equation. Model DE7 has a very similar behaviour to the other classes of 
models, in particular to the quintessence models. The linear overdensity parameter is smaller than the one for the 
$\Lambda$CDM model. When $c^{2}_{\rm eff}=0$, this model is almost identical to the $\Lambda$CDM model, while it 
differs substantially when $c^{2}_{\rm eff}=1$. All the models with effective sound speed $c^{2}_{\rm eff}=0$, 
converge very rapidly ($z\gtrsim 3$) to the $\Lambda$CDM model and hence to the EdS model, since DE becomes negligible 
at such high redshifts. Different again is the situation for the $c^{2}_{\rm eff}=1$ case, where the DE models recover 
the $\Lambda$CDM model at much higher redshifts (quintessence models), while phantom models reproduce the reference 
model very quickly. Models DE5 and DE7 with $c^{2}_{\rm eff}=1$ instead do not recover the $\Lambda$CDM model even at 
high redshifts. As said before, this is largely due to the additional source for the gravitational potential and 
largely independent of the background equation of state $w$, as shown in \cite{Pace2010}.

In Figure~\ref{fig:deltac_rot} we show results for $\delta_{\rm c}$ when the shear and rotation terms are taken into 
account for the DM Euler equation. As already shown and discussed in 
\cite{DelPopolo2013a,DelPopolo2013b,DelPopolo2013c}, the main effect appears at galactic scales 
($M\approx 10^{11}~M_{\odot}/h$). We verified that this is indeed the case also for clustering dark energy models, 
therefore we will limit ourselves to present our results for galactic scale objects. The differences between the 
spherical collapse model and the extended spherical collapse model become increasingly smaller with increasing mass, 
disappearing at cluster scales. Qualitatively, therefore, clustering and non-clustering dark energy models behave in 
the same way with respect to the mass dependence. We refer to the caption for line styles and colours of each model.

As expected, and in analogy with the extended spherical collapse model, when the influence of DE is only at the 
background level \citep{DelPopolo2013b}, the additional term opposes to the collapse, therefore the values for the 
linear overdensity parameter are higher than for the case in which these terms are neglected. Also in this case, 
quintessence models with $c^2_{\rm eff}=1$ differ more from the $\Lambda$CDM model than for the case with 
$c^2_{\rm eff}=0$. Phantom models are now very similar to the $\Lambda$CDM model, differently from before. Differences 
between the case with $c^2_{\rm eff}=0$ and $c^2_{\rm eff}=1$ are now negligible. Model DE7 behaves qualitatively as 
for the standard spherical collapse model. Also in this case all the models, except for the models DE5 and DE7 with 
$c^2_{\rm eff}=1$, recover the $\Lambda$CDM model at high redshifts. As shown in \cite{DelPopolo2013a,DelPopolo2013b}, 
in the ESCM major differences take place at low redshifts. We can therefore conclude that clustering DE models behave 
similarly to the non-clustering DE models when shear and rotations are included in the analysis.

However DE and its perturbations can also affect the virialization process of dark matter. A reference work focusing 
on this issue is \cite{Maor2005}. In this seminal work it was shown that a different result for the ratio between 
the virialization radius and the turn-around radius (the radius of maximum expansion) $y$ changes according to the 
recipe used, in particular if the dark energy takes part or not into the virialization process. Whatever is the 
correct formulation for the virialization process in clustering DE models, our ignorance on the exact value of $y$ 
will not qualitatively affect our discussion and conclusions, therefore for simplicity we will use $y=1/2$, as in the 
Einstein-de Sitter model \citep[see also the discussion in][]{Batista2013}. Since clustering dark energy does not 
alter the temporal evolution of the dark matter energy density parameter, we can still write 
$\Delta_{\rm V}=\zeta(x/y)^3$, where $\zeta$ represents the non-linear overdensity at turn-around, $x$ is the scale 
factor normalised at the turn-around scale factor. Our results for the (non-)rotating case are presented in 
Figure~(\ref{fig:deltav_norot})~\ref{fig:deltav_rot}.

As before, we limit ourselves to the study of the effects of the shear and rotation terms at galactic scales, since 
this is the mass scale where the effect is stronger. As for the $\delta_{\rm c}$ parameter, also in this case the DE 
models differ mostly from the reference model when the effective sound speed is of the order unity, while for 
$c^2_{\rm eff}=0$ the models are closer to the $\Lambda$CDM model. We also notice that, since at high redshifts the 
amount of dark energy is negligible, DE models recover the $\Lambda$CDM model. The model differing more is, once 
again, the DE5 with $c^2_{\rm eff}=1$ (see Figure~\ref{fig:deltac_norot}). Quintessence (phantom) models have lower 
(higher) values of $\Delta_{\rm V}$ with respect to the $\Lambda$CDM model. These results are qualitatively similar to 
what found in \cite{Pace2010}. Model DE7 behaves like the quintessence models having slightly smaller values for the 
case $c^2_{\rm eff}=0$.

We find qualitatively similar results in the ESCM (see Figure~\ref{fig:deltav_rot}). With respect to the usual case, 
we observe, as expected, that the virial overdensity is higher than for the usual spherical collapse model but the 
$\Lambda$CDM model is recovered at high redshifts. Once again major differences take place when $c^2_{\rm eff}=1$. We 
notice that our results are similar to what was found in \cite{DelPopolo2013b}. We can therefore conclude that 
clustering DE models behave qualitatively as non-clustering DE models in both the spherical and extended spherical 
collapse model. Shear and rotation terms only oppose to the collapse, without modifying it.

As said before, we have made the assumption that the shear and the rotation terms affect only dark matter. We 
performed a similar analysis relaxing this assumption and supposing that both dark matter and dark energy are 
influenced by these additional non-linear terms, then using a single equation for the velocity field, 
Equation~\ref{eqn:thetatp_alpha}. The results obtained are very similar to what presented here, therefore in the 
following, we will assume that DM and DE have a different peculiar velocity.

\subsection{Mass function}\label{sect:mf}
\begin{figure*}
 \centering
 \includegraphics[scale=0.45,angle=-90]{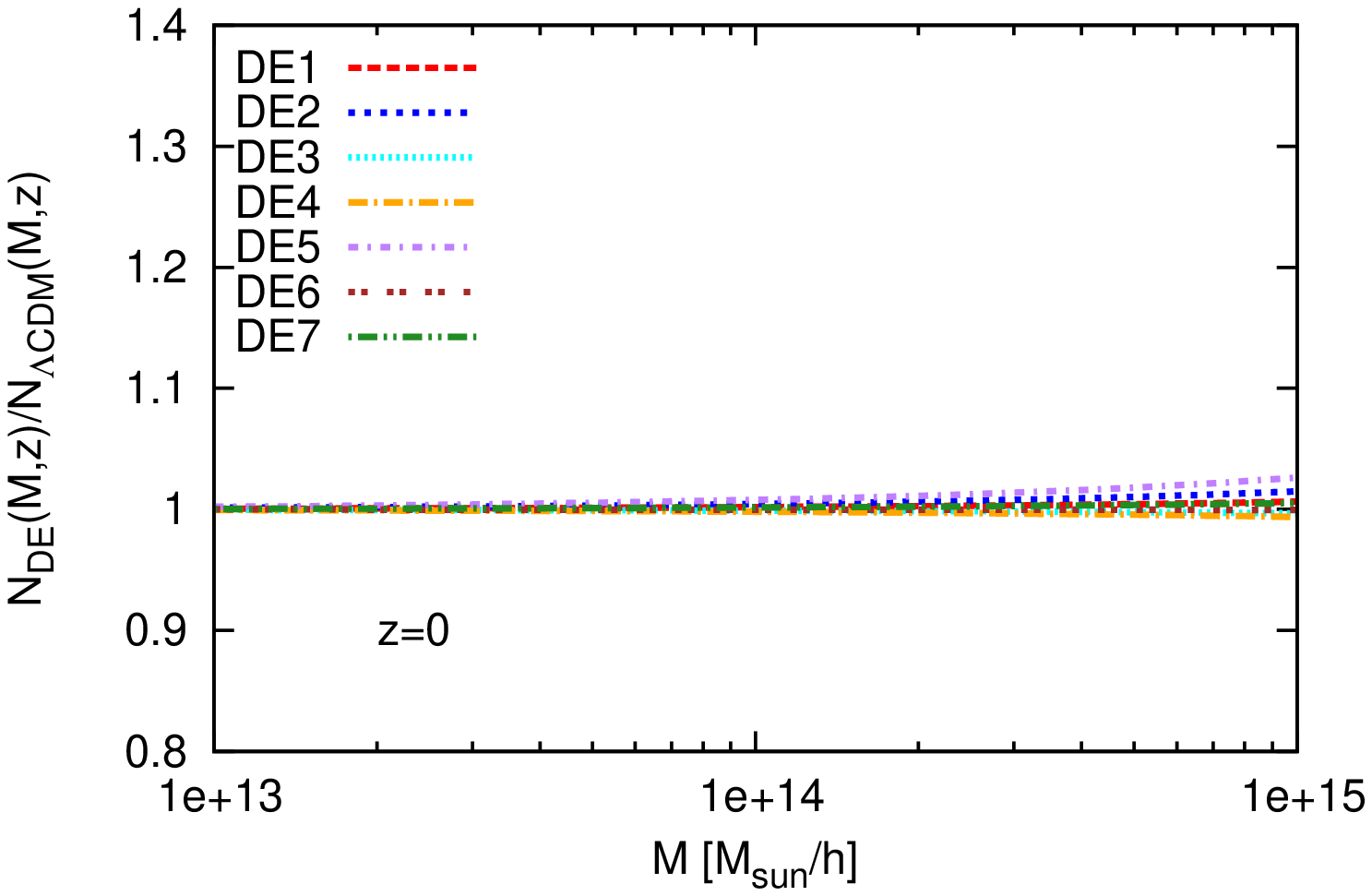}
 \includegraphics[scale=0.45,angle=-90]{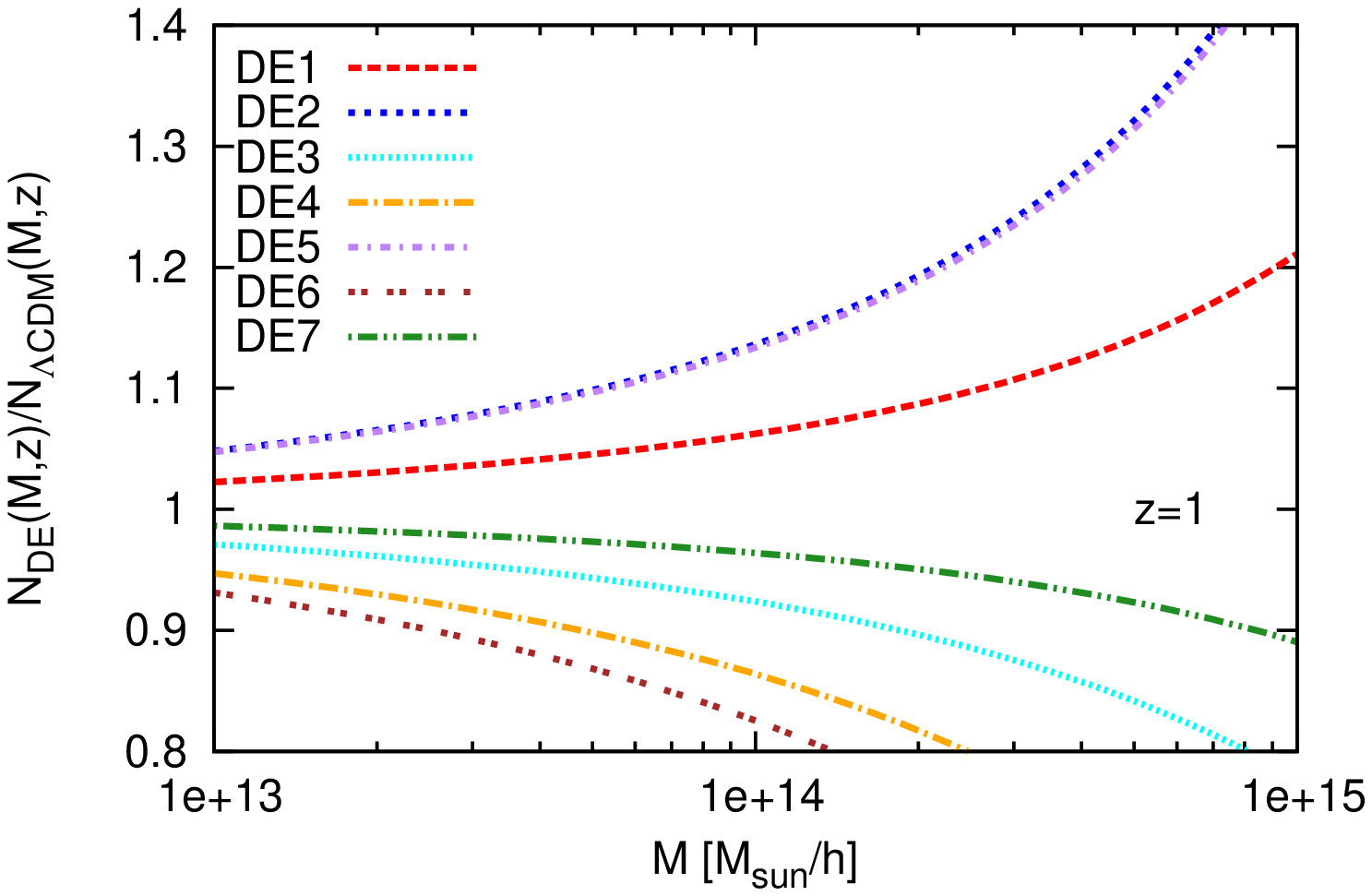}\hfill
 \includegraphics[scale=0.45,angle=-90]{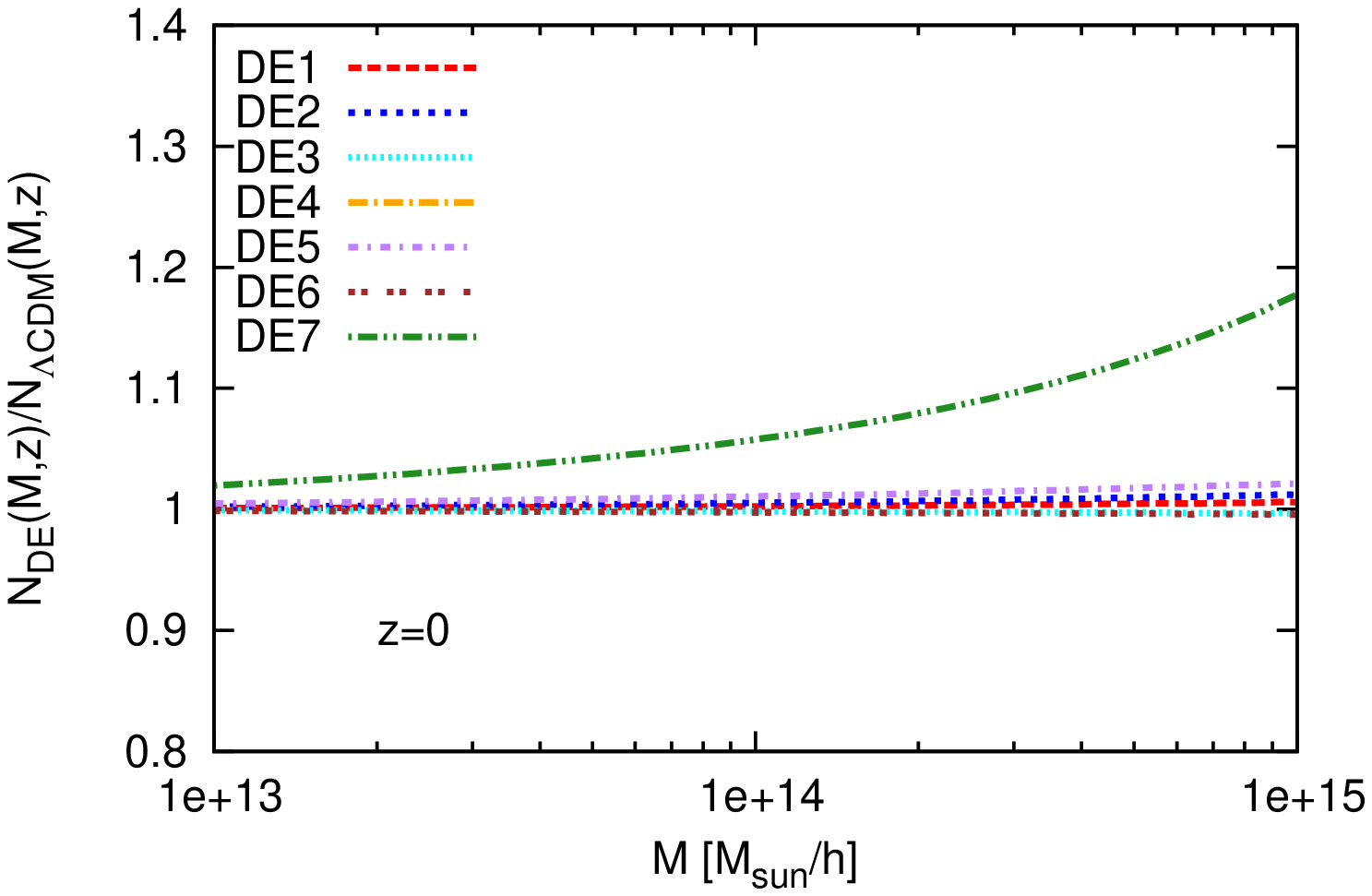}
 \includegraphics[scale=0.45,angle=-90]{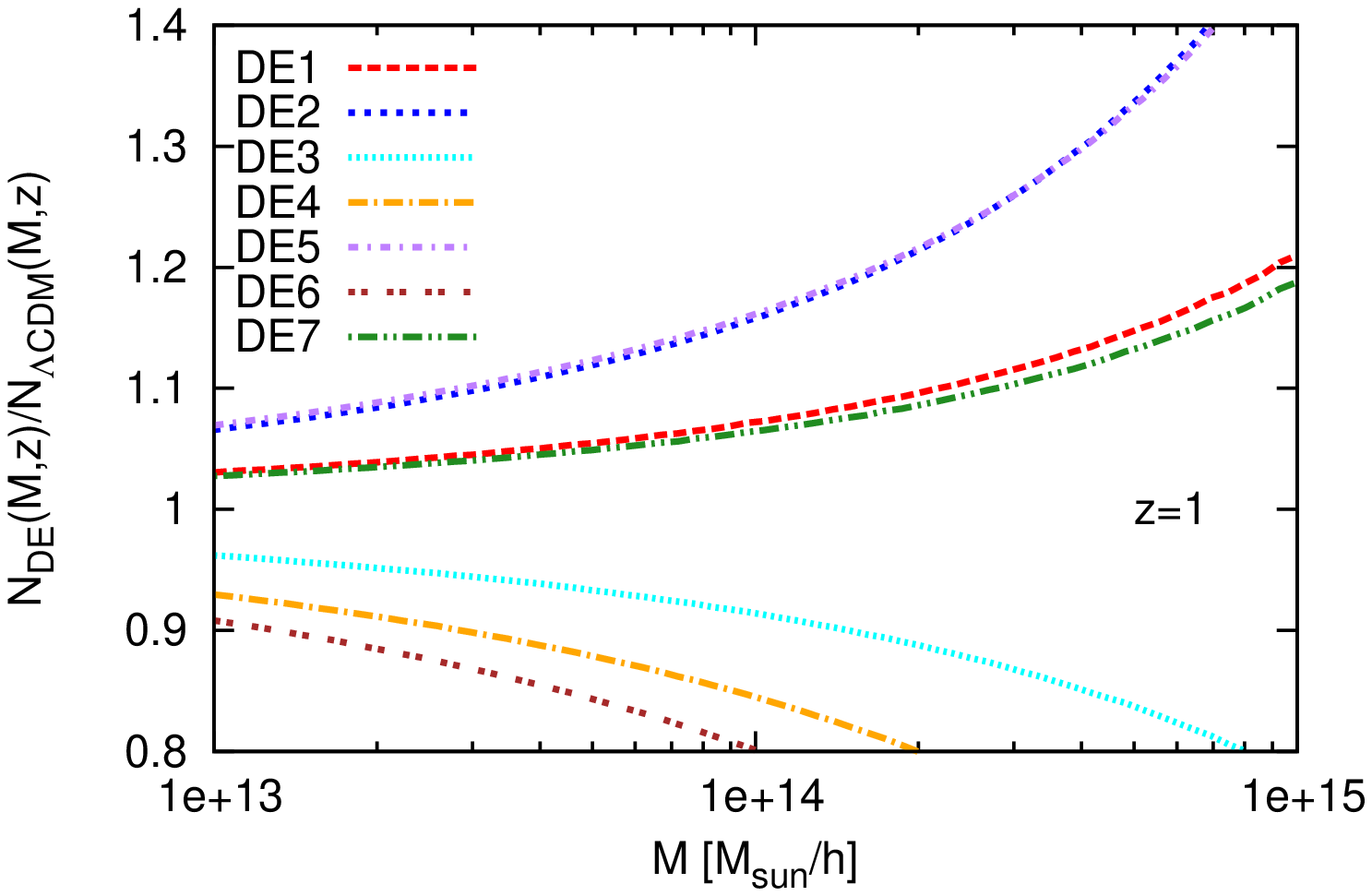}
 \caption{Ratio of the number of objects above a given mass $M$ for halos at $z=0$ (left panels) and $z=1$ (right 
panels) between the DE models and the $\Lambda$CDM model. The upper panels show ratios for the usual spherical 
collapse model while the bottom panels ratios for the extended spherical collapse model. The red dashed curve 
represents the DE1 model, the blue short-dashed curve the DE2 model, the cyan dotted curve the DE3 model, the yellow 
dot-dashed curve the DE4 model, the violet dot-short-dashed curve the DE5 model, the brown dot-dotted curve the DE6 
model and the green dashed-dot-dotted curve the DE7 model.}
 \label{fig:mf_rot}
\end{figure*}

Here we study the effect of the shear and rotation terms on the number counts of halos. We assume that the analytical 
formulation by Sheth \& Tormen \citep[][]{Sheth1999,Sheth2001,Sheth2002} is valid also for clustering dark energy 
models without any modification \citep[but see also][]{DelPopolo1999,DelPopolo2006}. The mass function critically 
depends on the linear overdensity parameter $\delta_{\rm c}$, the growth factor and on the linear power spectrum 
normalization $\sigma_8$. To properly evaluate the effects of the extra terms in the spherical collapse model 
formalism, we assume that all the models have the same $\sigma_8$ at $z=0$ and to highlight the effect we consider 
the number of objects at $z=0$ and $z=1$ above a given mass $M$ of the halo. We will assume as transfer function for 
the linear matter power spectrum the functional form given by \cite{Bardeen1986} and $\sigma_8=0.776$ as 
normalization of the power spectrum, in agreement with the most recent measurements 
\citep{Planck2013_XVI,Planck2013_XX}.\\
A comment is necessary at this point to explain the choice of the matter power spectrum normalization. In 
\cite{Batista2013} and in \cite{DelPopolo2013b} we had a different normalization for each model, such that all the 
models would have the same amplitude of perturbations at the CMB epoch ($z\approx 1100$). Here instead we enforce all 
the models to share the same normalization today. While a model dependant normalization should be in general 
preferred, here we want to isolate the effect of the shear and rotation terms and analyse their behaviour. If we 
would also have a different $\sigma_8$ for each model, a direct comparison between them would be more difficult.

In Figure~\ref{fig:mf_rot} we show the ratio of the number of objects above a given mass $M$ between the DE models and 
the $\Lambda$CDM model, restricting the analysis to the case $c^2_{\rm eff}=0$. In the upper (lower) panel we present 
results without (with) rotation and shear terms. Left (right) panels are for halos at $z=0$ ($z=1$). As expected, 
having the models the same normalization of the matter power spectrum, at $z=0$ the models have essentially the same 
number of objects, with very small differences for masses $M\approx 10^{15}$. In particular quintessence models (DE1, 
DE2, DE5) show a slight excess of structures, while phantom models (DE3, DE4, DE6) show a decrement in the number of 
structures. The model DE7 is at all effects identical to the model DE1. Differences grow with redshift and at $z=1$, 
they can be few tens of percent, keeping though the same qualitative behaviour. Models with highest differences are 
those with the equation-of-state parameter differing mostly from $w=-1$. Interestingly enough, the model DE7 shows now 
a decrement of $\simeq 10\%$ with respect to the $\Lambda$CDM model. With the inclusion of the shear and rotation 
terms, we see a behaviour qualitatively similar to the standard case, with major differences at higher redshifts (due 
to the time evolution of the dark energy), but with a smaller number of objects, at the level of percent for the 
phantom models, while quintessence models are largely unaffected. Results are consistent with the time evolution of 
the linear overdensity factor $\delta_{\rm c}$ (see Figures~\ref{fig:deltac_norot} and~\ref{fig:deltac_rot}). It is
interesting the case of the model DE7, which shows a very strong dependence on the inclusion of the non-linear terms 
already at very low redshifts, but its time-dependence is very weak. This is probably due to the change of regime 
between the phantom and the quintessence one. Note that ratios shown in the second raw, are taken with respect to the 
$\Lambda$CDM model evaluated in the ESCM.

Dark energy models with $c^2_{\rm eff}=1$ \citep[see also][]{DelPopolo2013c} deserve a comment. Due to the major 
differences between the linear overdensity parameter $\delta_{\rm c}$ of these models with respect to the $\Lambda$CDM 
one, we expect differences already at $z=0$ and they increase in value at $z=1$. Qualitatively, the same behaviour is 
nevertheless recovered, so models DE3, DE4 and DE6 (phantom models) show a decrement in the number of objects. 
Analogously to the case with $c^2_{\rm eff}=0$, the two additional nonlinear terms just slightly increase differences 
with respect to the standard case, in agreement with Figures~\ref{fig:deltac_norot} and~\ref{fig:deltac_rot}.

As explained before in Sect.~\ref{sect:escm_params}, the linear overdensity parameter $\delta_{\rm c}$ becomes 
mass dependant when shear and rotation are included. To evaluate the mass function when these additional non-linear 
terms are included, we therefore explicitly evaluate the contributions of the shear and rotation terms to 
$\delta_{\rm c}$ for each mass, so to have an exact evaluation of the mass function. This means that major differences 
will take place at small masses, while at cluster scales differences between the two different mass functions will be 
negligible. This is indeed the case in the lower panels of Fig.~\ref{fig:mf_rot}, with the exceptions of the model DE7 
as clarified above.

After establishing the effect of the shear and rotation terms on the mass function, we investigate deeper the effects 
of the clustering of the DE fluid. In this case, the total mass of the halo is affected by dark energy perturbations
\citep{Creminelli2010,Basse2011,Batista2013} and we need to take this into account evaluating the fraction of the 
halo mass given by the clustering of the dark energy. How and how much dark energy contributes to the halo mass 
depends on the virialization process, in particular whether dark energy virializes and on which time scale. If the 
halo mass is modified, then also the merging history \citep[see e.g.][]{Lacey1993} must reflect somehow this 
additional contribution. 
According to the equation-of-state parameter, DE can add or subtract mass to the total halo mass. An exact treatment 
of this problem must take into account the nature of the dark energy fluid and its exact virialization process. This 
is beyond the purpose of this work and we will use an approximate recipe, limiting ourselves to the case in which 
$c_{\rm eff}^2=0$ and we will assume that DE virializes with DM on the same time scale \citep[see][]{Batista2013}. In 
the following we will describe how to evaluate the fraction of DE with respect to the total mass of the halo 
\citep{Batista2013}.

As done in Section~\ref{sect:escm_params}, we will assume that $y=R_{\rm vir}/R_{\rm ta}=1/2$ as in the Einstein-de 
Sitter (EdS) universe. For this model, the virial overdensity $\Delta_{\rm V}$ can be evaluated analytically and in 
literature two different definitions are usually adopted. The most common one \citep{Wang1998}, evaluates it at the 
collapse redshift $z_{\rm c}$: 
$\Delta_{\rm V}(z_{\rm c})=\rho_{\rm m}(z_{\rm v})/\bar{\rho}_{\rm m}(z_{\rm c})\simeq 178$, where $z_{\rm v}$ is the 
virialization redshift. According to \cite{Lee2010} and \cite{Meyer2012}, it is more correct to evaluate it at the 
virialization redshift: $\Delta_{\rm V}(z_{\rm v})=\rho_{\rm m}(z_{\rm v})/\bar{\rho}_{\rm m}(z_{\rm v})\simeq 147$. 
These values will obviously change in the presence of DE and they depend on the properties of DE 
\citep{Lahav1991,Maor2005,Creminelli2010,Basse2011}.

The fraction ($\epsilon(z)$) of DE mass ($M_{\rm DE}$) with respect to the DM mass ($M_{\rm DM}$) is
\begin{equation}\label{eqn:eps}
 \epsilon(z)=\frac{M_{\rm DE}}{M_{\rm DM}}\;.
\end{equation}
We define the DM mass as
\begin{equation}\label{eqn:DM_mass}
 M_{\rm DM}=4\pi \int_0 ^{R_{\rm vir}} dR R^2 (\bar{\rho}_{\rm DM}+\delta\rho_{\rm DM})\;,
\end{equation}
and the DE mass as
\begin{equation}\label{eqn:DE_mass_pert}
 M_{\rm DE_{\rm P}}= 4 \pi \int_0 ^{R_{\rm vir}} dR R^2 \delta\rho_{\rm DE}(1+3c^2_{\rm eff})\;.
\end{equation}
We label the DE mass as $M_{\rm DE_{\rm P}}$ (see Equation~\ref{eqn:DE_mass_pert}) to indicate that in its definition 
we consider only the contribution coming from the perturbation.\\
If instead we consider also the background contribution, the mass definition becomes
\begin{equation}\label{eqn:DE_mass_tot}
M_{\rm DE_{\rm T}}= 4 \pi \int_0 ^{R_{vir}} dR R^2
\left[(1+3w)\bar{\rho}_{\rm DE}+(1+3c^2_{\rm eff})\delta\rho_{\rm DE}\right]\;,
\end{equation}
in analogy with the Poisson equation. In this case there will be also a contribution for the $\Lambda$CDM model. 
However, since the background term varies in time, regardless of the halo formation history, this contribution is not 
constant and should be interpreted just as a crude estimate of the background DE energy to the halo mass.

\begin{figure}
 \centering
 \includegraphics[scale=0.5,angle=-90]{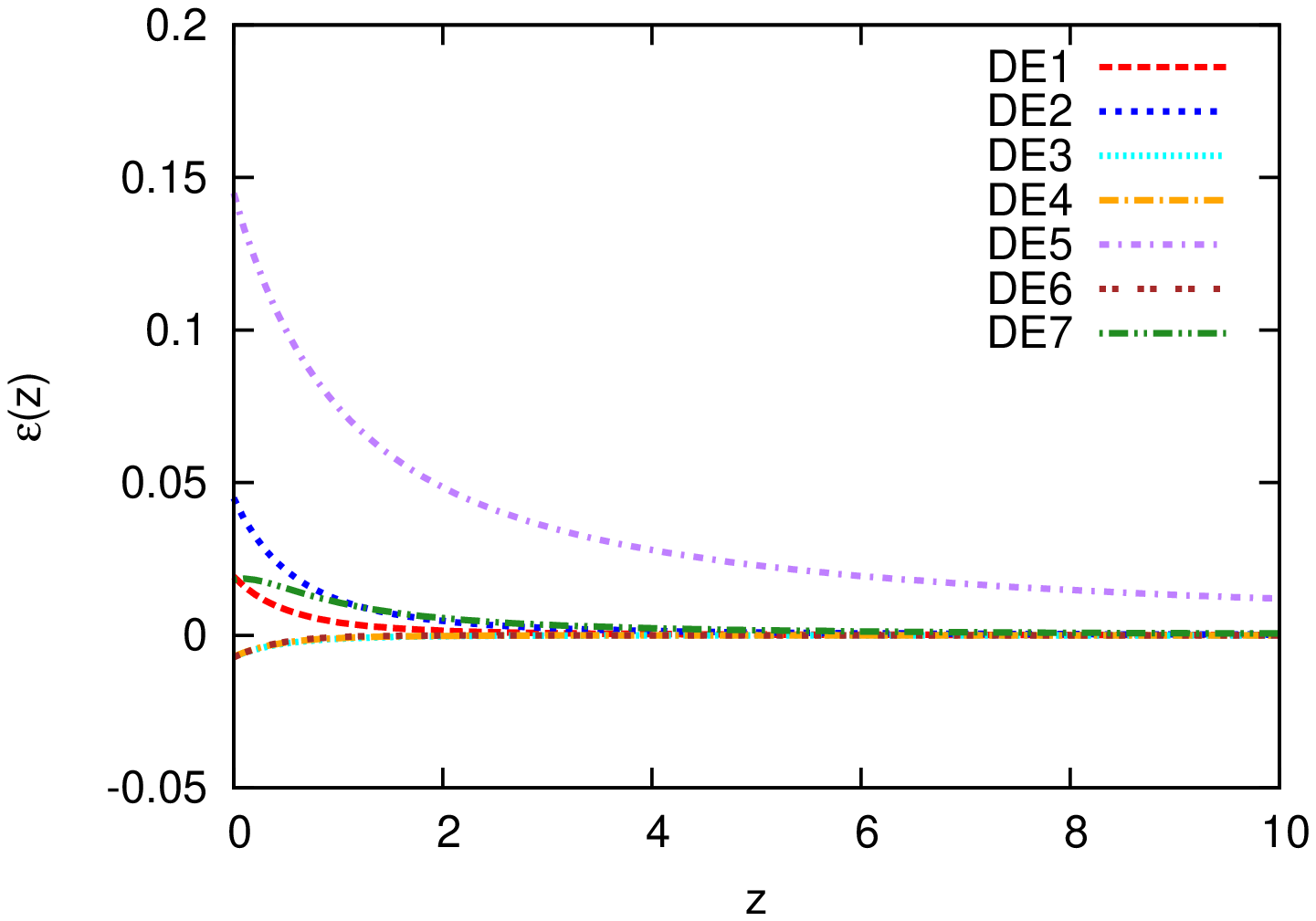}
 \includegraphics[scale=0.5,angle=-90]{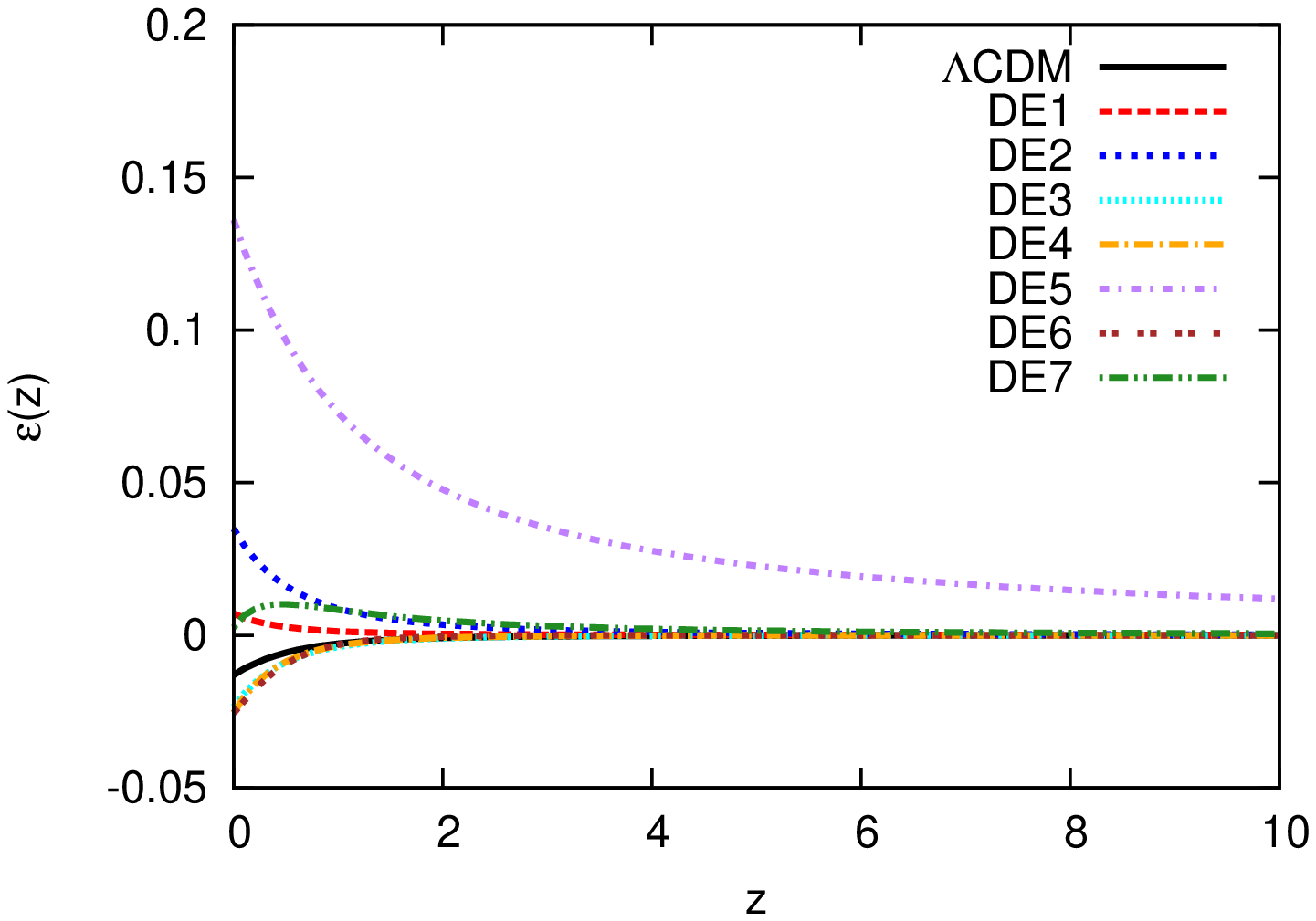}
 \caption{Fraction of the DE mass with respect to the DM mass according to the definition of 
Equations~\ref{eqn:DE_mass_pert} (upper panel) and~\ref{eqn:DE_mass_tot} (lower panel). The black solid curve shows 
the $\Lambda$CDM model, the red dashed curve the DE1 model, the blue short-dashed curve the DE2 model, the cyan dotted 
curve the DE3 model, the yellow dot-dashed curve the DE4 model, the violet dot-short-dashed curve the DE5 model, the 
brown dot-dotted curve the DE6 model and the green dashed-dot-dotted curve the DE7 model.}
 \label{fig:eps}
\end{figure}

In Figure~\ref{fig:eps} we show the fraction of the DE mass with respect to the DM mass according to the definition 
used in Equations~\ref{eqn:DE_mass_pert} (upper panel) and~\ref{eqn:DE_mass_tot} (lower panel) for the case 
$c^2_{\rm eff}=0$ only in the standard spherical collapse model. For a deeper discussion on the mass definition 
adopted see \cite{Batista2013}. We just show results for the standard spherical collapse model since rotation and 
shear have a negligible effect on $\epsilon(z)$. In particular for quintessence models the extra terms slightly 
reduce the DE contribution, while for the phantom models, being $\epsilon(z)$ negative and therefore subtracting mass 
to the halo, this function is slightly higher, or in absolute values, again slightly smaller. Same result for the 
barrier crossing model. The effect of the shear and rotation terms is of the order of tenth of percent.

\begin{figure*}
 \centering
 \includegraphics[scale=0.45,angle=-90]{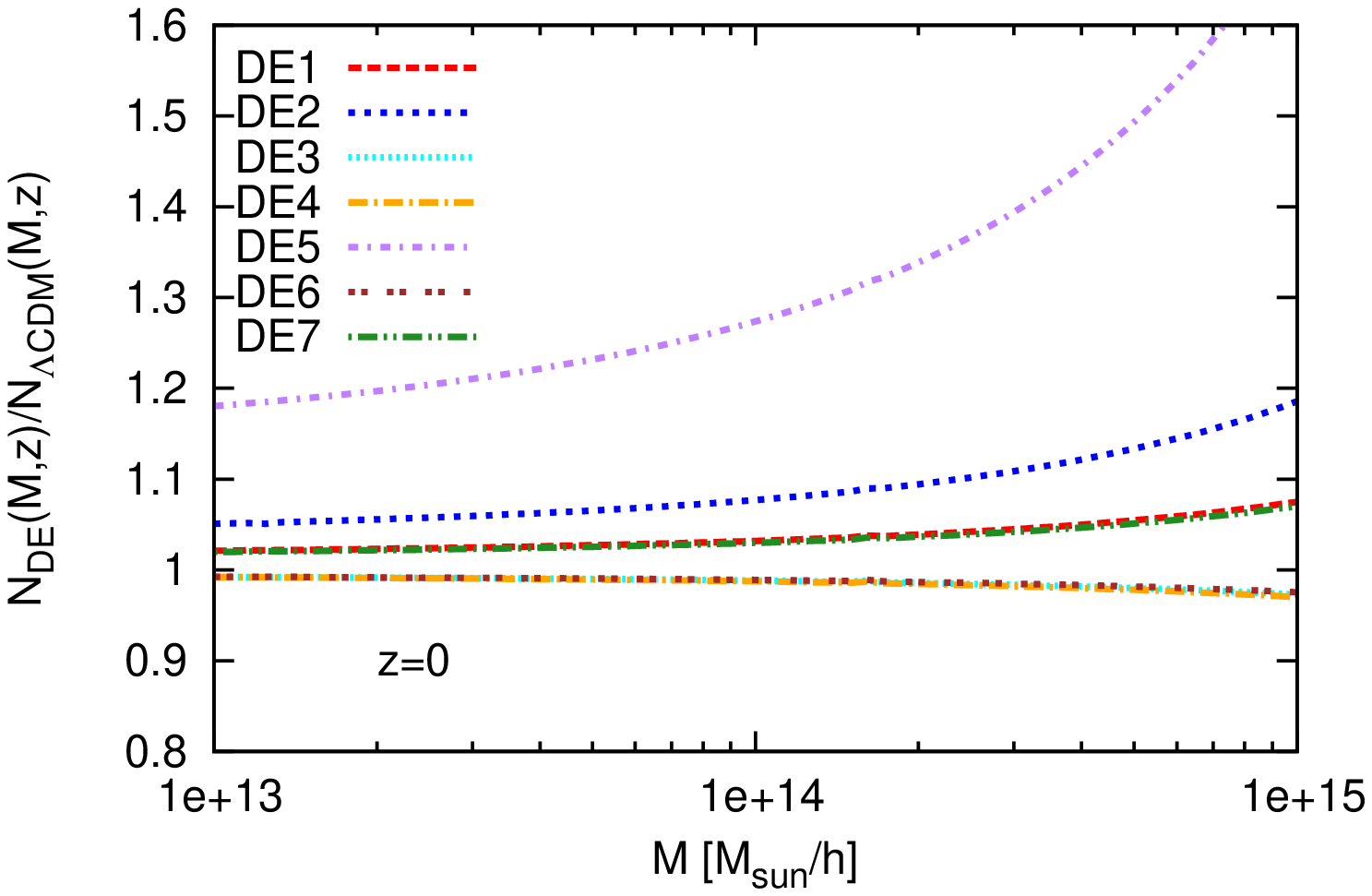}
 \includegraphics[scale=0.45,angle=-90]{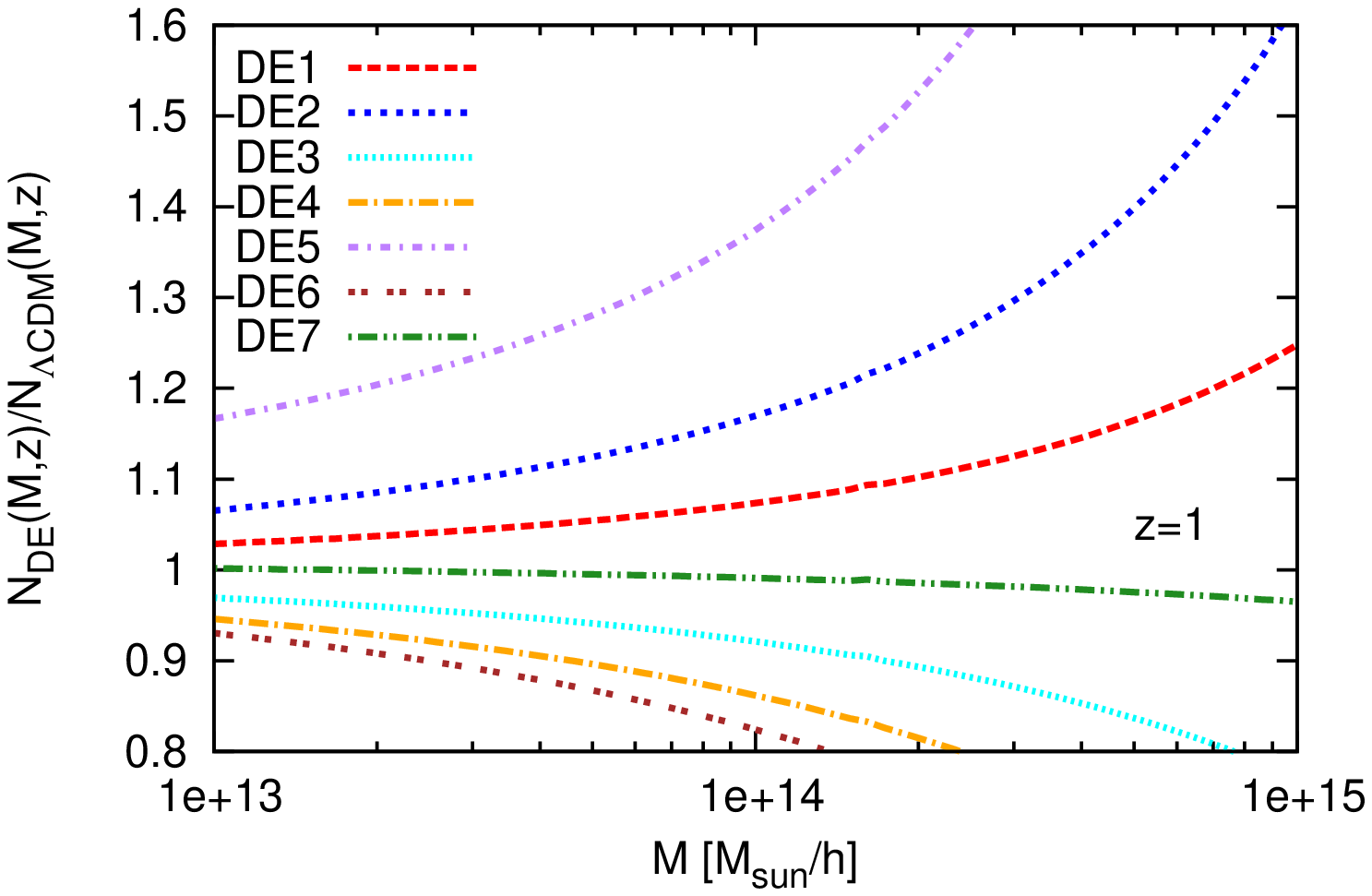}\hfill
 \includegraphics[scale=0.45,angle=-90]{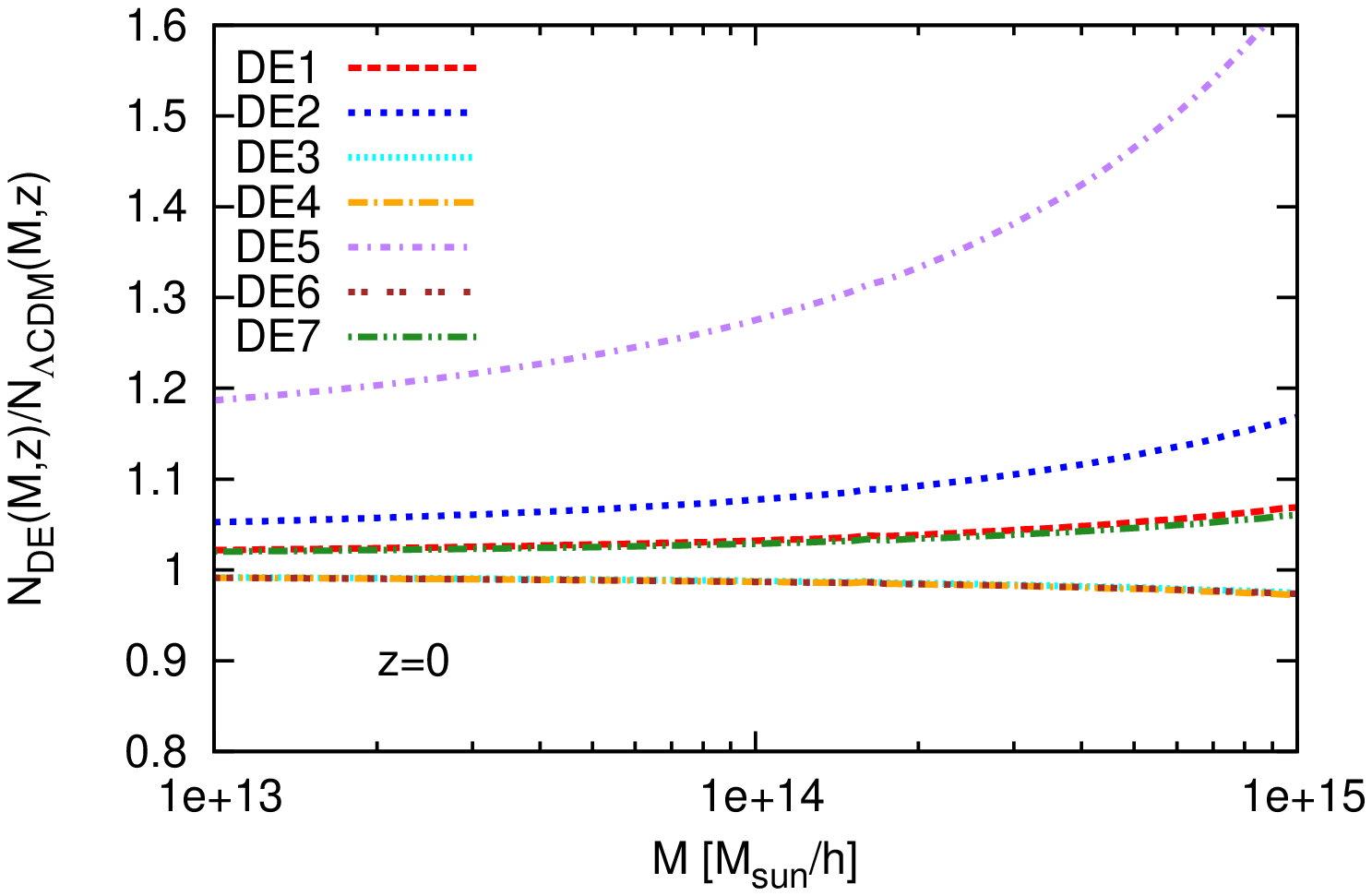}
 \includegraphics[scale=0.45,angle=-90]{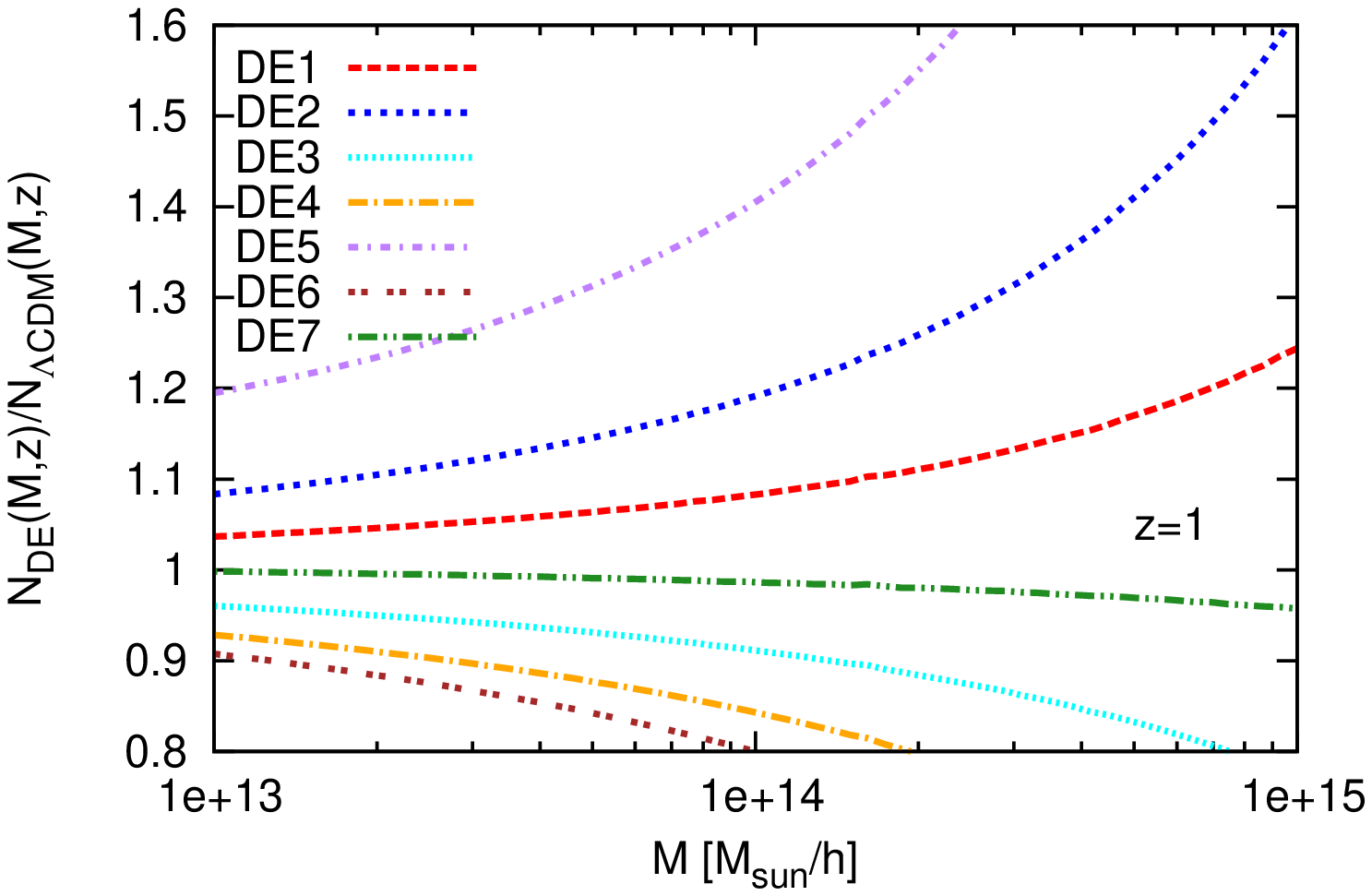}
 \caption{Ratio of the number of objects above a given mass $M$ for halos at $z=0$ (left panels) and $z=1$ (right 
panels) between the DE models and the $\Lambda$CDM model, using the mass definition in 
Equation~\ref{eqn:DE_mass_pert}. The upper panels show ratios for the usual spherical collapse model while the bottom 
panels show the ratios for the extended spherical collapse model. Line styles and colours are as in 
Figure~\ref{fig:mf_rot}.}
 \label{fig:mf_eps_rot}
\end{figure*}

As expected (see Figure~\ref{fig:eps}), quintessence models give a positive contribution to the total mass of the 
halo while phantom models subtract mass. Differences are of the order of the percent level, except for the model DE5, 
where differences are up to $\approx 14\%$. In agreement with \cite{Batista2013}, we also notice that the mass 
correction evaluated with Equation~\ref{eqn:DE_mass_tot}  is smaller than when only perturbations are taken into 
account. Major differences are at $z=0$ and become null at higher redshifts. This is expected, since $\epsilon(z)$ is 
significantly different from zero at low redshifts. Exception is once again model DE5. This is due to the fact that 
its equation of state is very different from $w=-1$. The inclusion of a mass correction term will affect the mass 
function and major differences will take place for $z\approx 0$, as we show in Figure~\ref{fig:mf_eps_rot}. 
Differences are again more pronounced for high masses where they can be up to 20\% while for low masses it is only of 
the order of 5\% at most. The hierarchy of the models, i.e., how much they are affected, directly reflects the values 
of the mass correction.

In Figure~\ref{fig:mf_eps_rot} we show the ratio of the mass function with the new mass definition, $M(1-\epsilon)$, 
where $\epsilon$ is given by Equation~(\ref{eqn:eps}). Since we only consider the contribution of the DE 
perturbations, the cosmological constant does not contribute to the total mass of the system. Quintessence (phantom) 
models have a lower (higher) number of objects at the low mass end of the mass function and a higher (lower) number of 
objects at the high mass tail. This can be easily explained taking into account the relative contribution of the DE 
component to the total mass of the halo (see upper panel of Figure~\ref{fig:eps}). A positive (negative) contribution 
to the total mass shifts the mass function towards lower (higher) values and, as consequence, we obtain a higher 
(lower) number of objects. At the high mass end, the contribution of the linear overdensity parameter $\delta_{\rm c}$ 
dominates, giving the opposite trend with respect to the low mass tail.\\
We can conclude that shear and rotation terms have in general a negligible contribution also when the mass definition 
in Equation~\ref{eqn:DE_mass_pert} is adopted.\\
Similar results apply for the case $c_{\rm eff}=1$, where we note that since perturbations in dark energy are 
negligible, so is the mass correction.

\section{Conclusions}\label{sect:conclusions}
In this work we studied the effect of the inclusion of the term $\sigma^2-\omega^2$. We analysed its impact in the 
framework of the spherical collapse model and in particular on the linear overdensity parameter $\delta_{\rm c}$ and 
on the virial overdensity $\Delta_{\rm V}$. The parameter $\delta_{\rm c}$ is one of the ingredients of the mass 
function and its variation reflects on the mass function and, as consequence, on the number of objects at a given 
redshift.

We consider dark matter and dark energy component as two fluids described by the respective equation of state and both 
of them can cluster. In particular, we relate the pressure perturbations to density perturbations for the dark energy 
component with the effective sound speed parameter $c_{\rm eff}^2$, that we assume to be constant and its values where 
fixed to $c_{\rm eff}^2=0$ and $c_{\rm eff}^2=1$, as currently done in literature.

The $\sigma^2-\omega^2$ term, being non-linear appears only in the non-linear equation describing the evolution of 
the peculiar velocity, therefore the growth factor is not affected. We made the assumption that only dark matter is 
affected by this additional non-linear term, but if we instead suppose that both DM and DE experience shear and 
rotation, we showed that results are largely unaffected, since for the models studied DE perturbations are 
subdominant.

We showed that the additional non-linear term opposes the collapse, as for the case in which the dark energy is only 
at the background level. Opposing the collapse, it makes such that both $\delta_{\rm c}$ and $\Delta_{\rm V}$ have a 
higher value with respect to the standard spherical collapse model. Increments in the linear overdensity parameter 
are of the order of 40\% for low masses, analogously to what found in \cite{DelPopolo2013b}, where the extended 
spherical collapse model was studied in non-clustering dark energy models. Quintessence models have always a lower 
value of $\delta_{\rm c}$, both in the standard and in the extended spherical collapse model. Phantom models instead 
present higher values, due to the faster expansion of the universe. A similar behaviour is found for the virial 
overdensity parameter $\Delta_{\rm V}$.

Differences in the spherical collapse model parameters reflect obviously in the mass function and in particular in the 
number of objects above a given mass (see Section~\ref{sect:mf}). To properly evaluate the effect of the additional 
term, we use the same normalization of the linear matter power spectrum for all the models. Moreover, considering the 
number of objects above a given mass, does not introduce any geometrical dependence on the results that will therefore 
depend only on the particular model considered (DE equation-of-state parameter and effective sound speed). Comparing 
results in the ESCM with the standard SCM, we notice that the differences are in general small, of the order of the 
percent for all the models considered in this work.

When dark energy clusters, following \cite{Batista2013}, we speculate that the halo mass can be modified by the 
inclusion of the dark energy perturbation into its definition. We therefore evaluate the correction to the halo mass 
and we found that this is generally of the order of few percent at low mass (but higher on cluster scales) and its 
sign (being positive or negative) depends on the equation of state of the dark energy component. The shear and 
rotation terms slightly modify this function, making it closer to zero when these terms are taken into account. Due 
to the small value of this correction factor, modifications in the number of objects is also small.

We can therefore conclude that effects of rotations in clustering dark energy models are modest and comparable to 
what found  in \cite{DelPopolo2013a,DelPopolo2013b} for non-clustering dark energy models. 
Hence we may also expect that the effect of clustering dark energy in more realistic models of structure formation 
can be well described by the usual spherical collapse model.

\section*{Acknowledgements}
FP is supported by STFC grant ST/H002774/1 and RCB thanks FAPERN for financial support. The authors would like to thank 
the anonymous referee for the valuable comments that improved our manuscript.

\bibliographystyle{mn2e}
\bibliography{RotationCDE.bbl}

\label{lastpage}

\end{document}